\documentclass[reprint,onecolumn,showpacs,prX]{revtex4}

\input epsf
\usepackage{graphicx}
\usepackage{epsfig}
\usepackage{mathrsfs}
\usepackage{amsfonts}
\usepackage{mathrsfs}
\usepackage{amsmath}
\usepackage{natbib}
\usepackage{epstopdf}
\usepackage{subfigure}
\usepackage{setspace}
\usepackage{color}
\begin{document}

\title{Fast nastic motion of plants and bio-inspired structures}

\author{Q. Guo$^{1,2}$, E. Dai$^{3}$, X. Han$^{4}$, S. Xie$^{5}$, E. Chao$^{3}$, Z. Chen$^{4}$}
\affiliation{ $^1$College of Materials Science and Engineering, FuJian University of Technology, Fuzhou 350108, China\\
$^2$Fujian Provincial Key Laboratory of Advanced Materials Processing and Application, Fuzhou 350108, China\\
$^3$Department of Biomedical Engineering, Washington University, St. Louis, MO 63130 USA\\
$^4$Thayer School of Engineering, Dartmouth College, Hanover,\\
New Hampshire, NH 03755, USA\\
$^5$Department of Energy, Environmental, and Chemical Engineering, Washington University, St. Louis, MO 63130 USA\\
\footnote{Electronic mail:  Zi.Chen@Dartmouth.edu. Q. Guo, E. Dai, and X. Han contributed equally.}
}

\date{\today}

\begin{abstract}
The capability to sense and respond to external mechanical stimuli at various timescales is essential to many physiological aspects in plants, including self-protection, intake of nutrients, and reproduction. Remarkably, some plants have evolved the ability to react to mechanical stimuli within a few seconds despite a lack of muscles and nerves. The fast movements of plants in response to mechanical stimuli have long captured the curiosity of scientists and engineers, but the mechanisms behind these rapid thigmonastic movements still are not understood completely. In this article, we provide an overview of such thigmonastic movements in several representative plants, including \emph{Dionaea}, \emph{Utricularia}, \emph{Aldrovanda}, \emph{Drosera}, and \emph{Mimosa}. In addition, we review a series of studies that present biomimetic structures inspired by fast moving plants. We hope that this article will shed light on the current status of research on the fast movements of plants and bioinspired structures and also promote interdisciplinary studies on both the fundamental mechanisms of plants' fast movements and biomimetic structures for engineering applications, such as artificial muscles, multi-stable structures, and bioinspired robots.
\end{abstract}


\maketitle


\section{Introduction}
When one thinks of fast movement, plants do not usually come to mind. The movements of most plants \cite{Darwin1880} typically involve a tropism, the growth or turning movement of a biological organism in response to a direction-dependent stimulus, and are normally only noticeable over the course of hours or days. Classic examples of tropisms include phototropic movements towards light, directional growth or reorientation in response to gravity (gravitropism), and touch sensitive growth of the roots (an example of thigmotropism) \cite{Darwin1880,Goriely1998, Gerbode2012, Wang2013,Silverberg2012}. In addition to tropisms, some carnivorous plants trap prey using nastic movements, or direction-independent, reversible orientational changes in response to direction-independent stimuli. Among these, of particular interest is the Venus flytrap (\emph{Dionaea}) which, upon consecutive triggering of the sensitive hairs on its namesake trap, can close a trap in less than a fraction of a second to capture its arthropod prey. This unusual plant, living in bogs with nutrient-deficient soil, was first discovered in 1760 by the North Carolina colonial governor Arthur Dobbs and was considered as ``the great wonder of the [plant] kingdom". Later, Charles Darwin \cite{Darwin} systematically investigated the mechanisms of the Venus flytrap's fast movement and called the plant ``one of the most wonderful in the world."

The movements of live plants are typcally classified into two categories, tropistic and nastic movements \cite{Burgert2009}. Tropistic movements refer to directional movements in response to external stimuli that have certain directionality (e.g., gravity, light, and mechanical touches). For example, the development of tendrils may depend on where and how they touch an object that acts as a support, so this type of movement is tropistic (termed thigmotropic movement or thigmotropism, where the prefix `thigmo' denotes touch responsive \cite{Scorza2011}). In contrast, nastic movements refer to direction-independent mechanical movements in response to external stimuli \cite{Braam2005}. For instance, the way that a \emph{Dionaea} trap closes usually does not depend on where and in what direction the sensitive hairs are touched, so it is an example of a nastic (or more specifically, thigmonastic) movement. On the other hand, researchers have also distinguished between active and passive systems among plant movements \cite{Hill1981}. Action potentials are typically involved in active movements, whereby the live tissues or cells can actively deform in response to the stimuli by the transport of ions and fluids or the change in membrane permeability, often assisted by mechanical principles such as buckling, swelling, or cavitation \cite{Burgert2009,Forterre2013}. In contrast, passive movements refer to configurational changes of dead tissues that occur in response to environmental changes \cite{Elbaum2007,Armon_2011,Guo2013}. Because of the vast amount of literature on plants' movement, we choose to focus primarily on fast, nastic movements and by ``fast movement" we refer to any movement that takes place in few seconds or less. 

Although it has been known for centuries that certain plants are capable of fast movement, the exact biomechanical, biochemical, and electrophysiological mechanisms which drive their fast movement have only begun to be comprehensively understood in recent years. For a more detailed overview of  existing knowledge of these mechanisms, the readers may refer to some recent review articles on fast moving plants \cite{Ueda2006,Fratl2009,Ellison2009,Martone2010,Joyeux2011b,Dumais2012,Zheng2013,Forterre2013,Poppinga2013,Moulia2013}. It is also worth noting that plant biomechanics and mechanobiology have garnered increasing attention from the scientific community, as they represent ``convergent paths to flourishing interdisciplinary research" \cite{Moulia2013}. In the past decades, the role of mechanical forces, both internal and external, has been considered critical in the shaping of biological shapes, such as in stem cell differentiation \cite{Discher2005}, embryonic morphogenesis \cite{Savin2011,Wyczalkowski2012,Kuhl2013,Gleghorn2013}, cancer cell migration \cite{Chihan2013}, plant morphogenesis \cite{Gerbode2012,Silverberg2012}, and touch-sensitive responses of plants \cite{Braam2005}. In this review, nevertheless, our focus is on the mechanically stimulated movement of plants.

The mechanical sensing, actuation, and movement in plants have since become sources of inspiration in biomimetic design, which have a wide range of engineering applications, including structural mechanics, biomedical engineering, and chemical engineering. In this article, we review the characteristics and mechanisms of several representative fast-moving plants and some recent development of engineering structures inspired by these plants. It is foreseeable that interdisciplinary collaborations between biologists, physicists, chemists, mathematicians, and engineers will bring forth new insights that ultimately lead to a unified picture of how these non-muscular systems operate. We hereby hope that our work can promote further interdisciplinary studies on the movements of plants as well as the advancement of bio-inspired and biomimetic technologies.

\section{Fast Movement of the Venus flytrap}
In this section, we examine the important structures and features of the Venus flytrap and review the biomechanical, chemical, and electrophysiological mechanisms behind its fast movement. Further study of these mechanisms will contribute to more comprehensive understanding of the nastic movements of plants.

\subsection{The structural features of the Venus flytrap}
\begin{figure}[th]
\begin{center}
\includegraphics[width=0.9\textwidth]{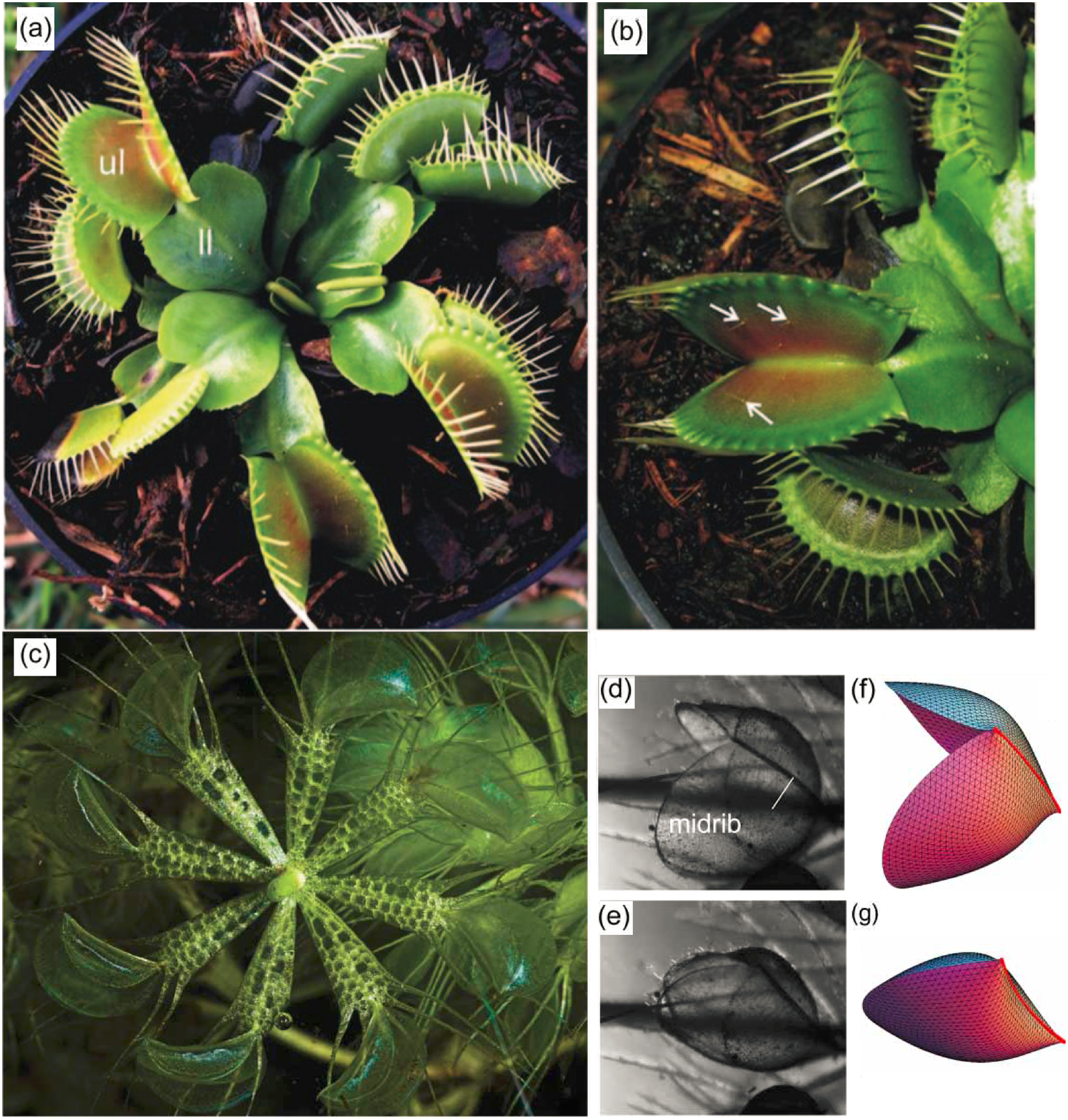}
\end{center}
\vspace*{8pt}
\caption{Images of \emph{Dionaea} ((a) and (b)) and \emph{Aldrovanda} ((c)-(g)). (a)
The modified leaves of \emph{Dionaea} are divided in two parts, the upper (ul) and the lower
leaf (ll). (b) The upper leaf has two lobes which center is brightly colored red and
contains three sensitive trigger hairs (arrows). The free edge of each lobe is lined with spine-like projections or cilia. (a) and (b) are adapted from Ref. \cite{Scorza2011}. (c) Photograph of a node of an \emph{Aldrovanda vesiculosa} whorl with eight leaves (photograph \copyright Dr. Barry Rice, Sierra College, Rocklin, CA, USA). (d, e) Microphotographs of a trap in open and closed configurations, respectively. The oval lobes are approximately 5 mm long. (f, g) Models of the trap in open and closed configurations, respectively. (d)-(g) are adapted from \cite{Poppinga2011}.} \label{flytrap}
\end{figure}

The most spectacular feature of the Venus flytrap, is the ``open mouth", a pair of leaves that traps arthropods to nourish the plant. As shown in Figure \ref{flytrap}(a), the modified leaves of the Venus flytrap have two parts, the upper and lower leaf. The lower leaf, also known as the petiolus, has an expanded leaf-like structure \cite{Volkov2008}. The upper leaf has three main features: a shell-like ``open mouth" composed of a pair of symmetrical lobes, three to five (typically three) sensitive hairs (see Figure \ref{flytrap}(b)) on the inner side of each lobe, and cilia, the numerous interlocked ``teeth" located on the border of each lobe. When an external force is applied to the trigger hairs, the trap swiftly closes. Usually, two consecutive touches are needed to trigger this rapid closure, a mechanism that helps ensure the plant does not waste energy on capturing non-prey. This movement can also be triggered by other external stimuli, such as electricity or heat \cite{Brown1910, Volkov2007, Volkov2011}. In addition, upon mechanical stimulation, touch receptors, the small trichomes located on the outer leaves and stems of the Venus flytrap, generate action potentials which can sensitize the trigger hairs for contact \cite{DiPalma}.

\subsection{Biomechanics of fast movements in the Venus flytrap}
The physical mechanisms behind the trap of the Venus flytrap are highly complex and remain incompletely understood. Through systematic observation and experimentation, Darwin \cite{Darwin} discovered that the inner layers of the Venus flytrap's lobes contract during closure, while the entire leaf's shape changes from convex to concave. Following these observations, researchers have proposed that the trap is mechanically bistable, with two stable states, i.e., the open and closed configurations, and an intermediate state \cite{Forterre, Volkov2008, Yang2010}.

\begin{figure}[th]
\begin{center}
\includegraphics[width=0.9\textwidth]{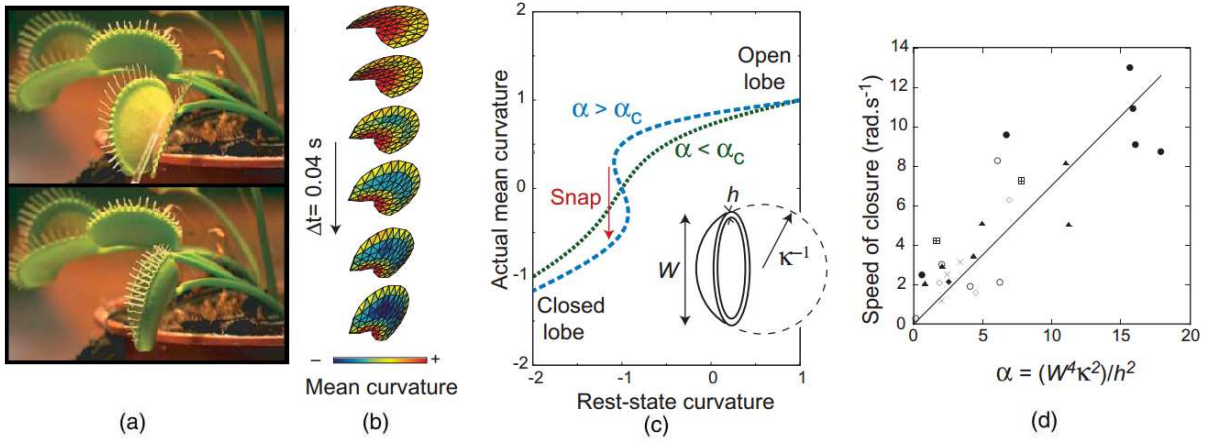}
\end{center}
\vspace*{8pt}
\caption{Snap-buckling of the Venus flytrap (\emph{Dionaea muscipula}). (a) Venus flytrap in its open and closed states, shown in the upper and lower panel, respectively.  (b) Computational reconstruction of trap lobe during closure from experimental data. (c) The mean curvature as a function of the dimensionless parameter $\alpha = W^4 \kappa^2/h^2$. For $\alpha < \alpha_c$. No multistability is predicted when $\alpha < \alpha_c$. When $\alpha > \alpha_c$, the system goes through a region featuring multistability where a ``snap-through" motion occurs.
(d) Average speed of trap closure as a function of $\alpha$. Adapted from Ref. \cite{Forterre} and Ref. \cite{Forterre2013}.}\label{flytrapNature}
\end{figure}

Following Darwin, several additional explanations have been proposed for the underlying mechanisms behind this trap closure. Brown \cite{Brown} proposed that trap closure is caused by the expansion of the outside of the lobes, and furthermore, that the opposite expansion of the inner surface of the lobes drives the reopening process of the Venus flytrap. Inspired by Darwin's hypotheses, Forterre \emph{et al.} used modern experimental technology to examine the biomechanics of leaf closure \cite{Forterre}. Using elasticity theory, they proposed that a dimensionless parameter, $\alpha = W^4 \kappa^2/h^2$ (where $W$ denotes the size of the lobe, $h$ is the thickness, and $\kappa$ is the curvature), controls bistability (see Figure \ref{flytrapNature}(a)-(c)). To test this hypothesis, Forterre and coworkers drew arrays of sub-millimetric ultraviolet-fluorescent dots on the surface of the leaves and irradiated the plant to fluoresce the dots, with the movements of these dots recorded by high speed camera during trap closure. From these experiments, the researchers inferred that the main source of the fast and dramatic closure of the trap is the bistable, doubly curved structure of the leaves, which snap-buckle to reverse their Gaussian curvature upon closing. Moreover, the sudden change of intrinsic curvature along the direction perpendicular to the midrib was determined to be the main source of the driving force behind trap closure, verified by measuring the strain field upon closure. The experiments also showed that the speed of the lobe during closure increases as a function of $\alpha$ (Figure \ref{flytrapNature}(d)).

While Forterre and coworkers demonstrated that the fast motion of the Venus flytrap involves a snap-buckling instability of the shell-like geometry of the lobes, the cell and tissue scale sources of this active movement remain poorly understood. To address these concerns, Colombani and Forterre \cite{ColombaniForterre} studied the cell scale \emph{in vivo} measurements of the pressure and other poroelastic properties using a microfluidic pressure probe. Their initial results shed light on the time course of cellular pressure and volume change during plant movement, as well as membrane permeability and Young's modulus.  The determined poroelastic time of 20-150s, much too large to account for rapid closure of the trap, suggests that some other cellular property (e.g., pre-stress of mesophyll cells) rather than active, osmotically driven water transport between cells could be the the cell scale source of rapid trap closure. Additional studies are necessary to fully understand the exact nature of this source.

In addition, Markin \emph{et al.} \cite{Markin} proposed a hydroelastic curvature model to interpret the mechanism of the active movement of the Venus Flytrap. Here, based on the assumption that the trap possesses curvature elasticity and consists of outer and inner hydraulic layers where different hydrostatic pressures can build up, the natural principal curvatures are determined by the hydraulic states of the two layers of the plant, which are in turn defined by their differing hydrostatic pressures. Similar to a bimetallic couple, this so-called bilayer couple can change shape rapidly in response to triggering of the sensitive hairs of the plant. This change is thought to occur through transportation of water via aquaporins between the two layers. Yang \emph{et al.} \cite{Yang2010} further developed a mathematical model to simulate the closing and opening dynamics of the Venus flytrap from the nonlinear dynamics and control perspective. In a separate study, Joyeux \cite{Joyeux2013} presented an elasticity model of the Venus flytrap by accounting for the anisotropic nature of the strain field of the trap and assuming that the open geometry is the minimum energy state, which embodies improvement over the previous model \cite{Joyeux2011} developed by Joyeux and coworkers. More recently, Pandolfi \emph{et al}. \cite{Pandolfi2014} showed that this movement is intimately dependent on gravity, which has not been considered before; however, the detailed biophysical mechanism remains inconclusively explained and requires further examination.

\subsection{Different ways of triggering the Venus flytrap}
The Venus flytrap can be triggered in a number of ways. In nature, the trigger mechanism is mechanical stimulation, in which two consecutive touches on the sensitive hairs within around 30 seconds lead to closure. These touches can be on the same hair or on separate hairs. Each touch generates an electrical signal, which transmits between the lobes and the midrib of the trap. In addition to the typical two touches, a single sustained touch to one trigger hair can also close the trap, if two electrical signals are generated about two seconds apart in the same manner \cite{Volkov2009}. Furthermore, there exists a temperature dependence of mechanically stimulated closure, with evidence suggesting that higher temperatures result in a greater sensitivity to mechanical stimulation \cite{Brown1910}.

External electrical stimulation, involving the transmission of electrical charge between a lobe and the midrib by means of electrodes and a charged capacitor can also induce closure. At room temperature, a total charge of about 14 $\mu$C \cite{Volkov2007}, applied at once or in smaller increments, also causes the Venus flytrap to close, though the exact amount of charge required increases with trap size and decreases at higher temperatures. Closure by repetitive application of smaller charges demonstrates electrical memory in the Venus flytrap (see the section \textit{Electrophysiology behind the Venus flytrap's fast movement}), subject to a time interval similar to that associated with mechanical stimulation \cite{Volkov2008}.

The Venus flytrap also closes in response to optical stimulation in the mid-infrared region. The snapping mechanism can be triggered with 7.35 $\mu$m radiation from a continuous-wave laser directed at the interior of a Venus flytrap. Furthermore, electrical signals are found to propagate in the Venus flytrap concurrently with each optical stimulation, in the same manner observed in mechanical stimulation \cite{Eisen}. Also noteworthy are recent experiments involving chemical stimuli of trap closure. Volkov \emph{et al.} \cite{Volkov2014} showed that application of 10 $\mu$L of ether or chloroform on the midrib induces slow (10s) closure of the trap, and similarly, that application of 10 $\mu$L of chloroform directly on the sensitive hairs can induce closure as fast as mechanical stimulation of the hairs.

\subsection{Biochemistry behind the Venus flytrap's fast movement}
A number of plant hormones have been identified and linked to several key properties of the Venus flytrap and its prey-catching and digesting mechanisms. In plants, lipid-based hormone signals known as jasmonates regulate defense mechanisms in response to abiotic and biotic stresses. In contrast, these jasmonates play a role in nutrion retrieval for the Venus flytrap: the phytohormone jasmonic acid and coronatine (the molecular mimic of the isoleucine of jasmonic acid) are jasmonates involved in the secretion of digestive enzymes. Perez \emph{et al.} demonstrated that spray-application of these jasmonates on just one leaf can trigger secretion of digestive enzymes throughout a Venus flytrap without any mechanical stimulation, indicating that these jasmonates are systemic in nature \cite{Perez}. In addition, the researchers showed that jasmonates are involved in the trap closing mechanism. They also showed that isolated traps from \emph{D. muscipula} can be induced to close when aqueous-phase extracts containing 12-oxo-phytodienoic acid are adsorbed on the leaves. In these experiments, the total quantity of bioactive substance, as opposed to the concentration, influences closure. One hypothesis suggests that this bioactive trap-closing chemical may be secreted stepwise in response to each action potential, inducing trap-leaf closure past a specific threshold \cite{Ueda2010,Volkov2007}. This hypothesis would thereby link the observed memory effect associated with the accumulation of stimuli to the accumulation of this chemical.

Abscisic acid (ABA) is also involved in the trap closing mechanism. ABA is a plant hormone that activates anion channels to close stomata, reducing transpiration. ABA is produced in plants in response to low water availability. In \emph{D. muscipula}, ABA has been shown to lower trap sensitivity to mechanical stimuli. Compared to well-watered Venus flytraps, both ABA-sprayed and water-deficient traps require additional mechanical stimulation to trigger fast trap closure \cite{Perez}.

\subsection{Electrophysiology behind the Venus flytrap's fast movement}
The Venus flytrap's rapid closure is typically triggered by consecutive mechanical stimuli within a window of 30 seconds on the one or multiple sensitive hairs. It is known that such mechanical stimulation activates certain ion channels to generate both a receptor potential and an action potential, which may be responsible for the opening of fluid channels between different hydraulic layers. Nevertheless, two consecutive stimuli only put the trap into a semi-closed state. Without any further stimulations, the trap will gradually reopen in a few hours. In the case where a trapped insect keeps struggling and stimulating the sensitive hairs, the trap will close tighter and tighter until it is fully sealed, upon which the digestive fluids are secreted to help digest the nutrients.

As mentioned before, an alternative way to trigger closure is by applying an electrical pulse above a threshold value between the upper layer of one lobe and the midrib. In such an experiment, the polarity is critical because an electrical pulse with inverted polarity does not trigger trap closure \cite{Markin}. Whether triggered mechanically or electrically, the rapid closure of the Venus flytrap always follows the generation of receptor and action potentials \cite{Volkov2007}. There are usually three phases during closure: a mechanically silent phase (no noticeable movement), an accelerating movement phase, and a fast movement phase \cite{Volkov2007, Zheng2013}. Volkov and co-workers \cite{Volkov2008} employed a high-speed data acquisition device to record the duration and amplitude of the action potentials and showed that the duration of an action potential in the Venus flytrap is typically about 1.5 milliseconds. Furthermore, the researchers \cite{Volkov2008} used ion channel uncouplers and blockers to explore the mechanisms of different closure phases. They found that both ion channel uncouplers and blockers can increase the time delay of trap closure dramatically while decreasing the speed of closure.

It is also interesting to note that the Venus flytrap exhibits short-term electrical memory \cite{Volkov2009}, i.e., separate electrical pulses, each of which is less than the threshold and within a time window of no more than 50 seconds, can initiate closure as long as the sum of electrical stimulus exceeds the threshold value. Volkov and coworkers demonstrated that this stimuli threshold is temperature dependent, with a charge threshold between 8 to 9 $\mu C$ (with a larger electrical threshold corresponding to larger traps) at room temperature compared to only 4.1 $\mu C$  at 28 to 36 degree Celsius \cite{Volkov2009}. This electrical memory was also demonstrated to be time dependent through application of below-threshold electric shocks separated by a 15 second interval, with decreased number of below-threshold electric shocks required for closure of the trap with increasing current of each shock. Researchers have also begun to discover the biomolecular components underlying this electrical memory. Ueda and co-workers \cite{Ueda2010} demonstrated that there exists a threshold of accumulated bio-metabolite for stimulating trap closure. Using bioassays to separate \emph{Dionaea} extracts, a bioactive polysaccharide was identified that is capable of triggering trap closure in absence of any external mechanical or electrical stimuli. The molecular electronics and different electrical characteristics of the complete hunting cycle (the open state, closed state, locked state, constriction and digestion state, and semi-open state) were further examined \cite{Volkov2009b,Volkov2011}. Putting the results from these studies together, it is promising that a more comprehensive understanding of the biological electronic system and the associated ``memory" and actuation mechanisms in plants such as \emph{Dionaea} might be achieved. Furthermore, while there are clear parallels to be drawn between the electrical signals in plants and animals, additional studies are required to elucidate the unique physiological origins and functions of these signals in plants \cite{Rainier}.

\section{Comparative biomechanics on active plant traps}
The mechanics of movements in \emph{D. muscipula} differ from those in the underwater carnivorous traps \emph{Aldrovanda vesiculosa} and \emph{Utricularia}. Poppinga and Joyeux compared the biomechanical mechanisms of \emph{Aldrovanda} and \emph{Dionaea} by modeling the \emph{Aldrovandas} trap as a pair of thin elastic shells hinged to the midrib \cite{Poppinga2011} (see Figure \ref{flytrap}(c)-(g)). The midrib connecting the two lobes can bend inward in the closed configuration. This study indicates that the \emph{Aldrovanda}'s trap closure is due to swelling and shrinking \cite{Skotheim2005} of tissues in or around the midrib, rather than mechanical buckling that takes place in \emph{Dionaea muscipula}. Moreover, the model suggests that this mechanism is the source of the rapid speed of trap closure in \emph{Aldrovanda}, i.e., a large opening or closing movement of the trap can result from a relatively small bending deformation of the midrib, in contrast with the snap-buckling of convex to concave curvature seen in the lobes of \emph {Dionaea}. More recently, Joyeux \cite{Joyeux2013} proposed a precise mechanism for the action of motor cells surrounding the midrib of a \emph{Aldrovanda} trap that could lead to closing and opening of the trap. This new model confirms that the reversible movements of \emph{Aldrovanda} trap can be controlled by the swelling and shrinking of the motor cells near the midrib, in contrast to the irreversible mechanism operating in \emph{Dionaea} traps where mechanical buckling plays a dominant role.

\begin{figure}[th]
\centering
\includegraphics[width=0.9\textwidth]{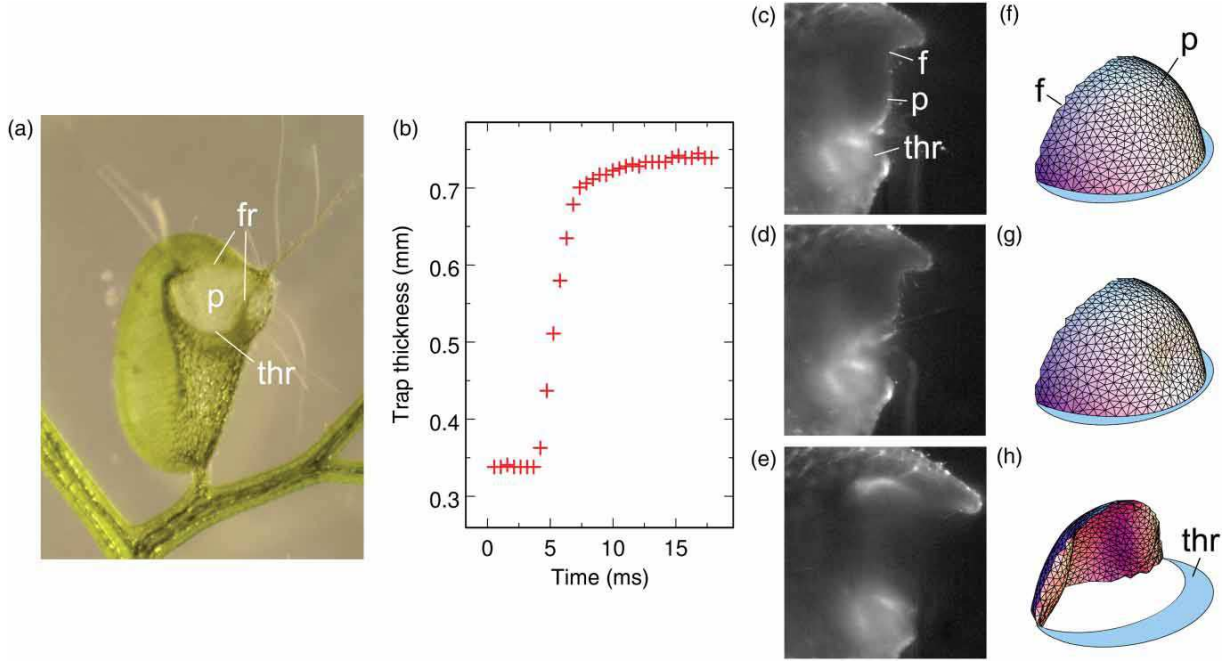}
\caption{
(a) \emph{Utricularia inflata} in untriggered, set configuration (adapted from Ref. \cite{Vincent1}). Note that the trap, approximately 3mm wide,  has concave curvature of its walls due to the negative internal pressure in set configuration. (b) Experimentally determined time course behavior of trap thickness during trap closure following mechanical stimulation of one of the trigger hairs. (c, d, e) Time course behavior of door opening of \emph{Utricularia australis} trap at 0, 7.6, and 8.6 ms following mechanical stimulation of trigger hairs, respectively. (f, g, h) Computational reconstruction of the trap door in set configuration (f), at the initial stages of buckling (g), and in the process of curvature inversion (h) door in set condition, at the onset of buckling, and while swinging open, respectively. (b)-(h) are adapted from Ref. \cite{Joyeux2011}. `p'  and `f' denote the panel and the frame of the door, respectively; `thr' denotes the threshold (following the conventions in Ref. \cite{Joyeux2011b}). \label{bladderwort}}
\end{figure}

Bladderwort (\emph{Utricularia}) traps also employ a mechanical buckling mechanism which differs from that of \emph{Dionaea} \cite{Vincent1}. \emph{Utricularia}'s suction traps are among the most complicated of all active traps in the plant kingdom \cite{Joyeux2011}. The trap is composed of a hollow trap body, a ``trap door" (which grants entrance to the trap body's interior), and four outward facing trigger hairs located on the trap door \cite{Joyeux2011}. During the trap-setting phase, active cells pump water out of the trap body through two-armed glands, termed bifids, creating a negative pressure differential up to -16 kPa \cite{Sydenham1985, Singh2011} between the interior and exterior of the trap body. As a result, the thin walls of \emph{Utricularia} become concave (modeled as spherical) with stored elastic energy in the doubly-curved shell of the trap body. Joyeux \emph{et al.} \cite{Joyeux2011} modeled this buckling scenario like a trap door. The researchers also showed that this buckling can also be initiated by a local deformation of the shell from mechanical stimulation of the trigger hair. Figure \ref{bladderwort} further suggests that the swinging motion of the panel starts with local curvature inversion, and that the ``door" acts like a flexible valve that flips under external pressure (and changes curvature from concave to convex)\cite{Joyeux2011}. Additionally, \emph{Utricularia} traps fire not only upon mechanical stimulation of the trigger hairs, but also spontaneously to capture small organisms and detritus (e.g., phytoplankton or bacteria) which are too small to mechanically trigger the trap \cite{Vincent2}. This spontaneous action occurs through mechanical oscillations of regular pumping and mechanical buckling of the trap door. The underlying mechanisms of the spontaneous firing could inspire biomimetic design of engineering structures that are capable of autonomous elastic deformation.\\

Furthermore, Llorens and co-workers developed a model to simulate the dynamical process of \emph{Utricularia}'s trap closure by considering hydrodynamics, buckling of the door, and deformation of the trap body. This comprehensive treatment can predict the main mechanical features of the \emph{Utricularia}'s trap closure, namely triggered closure and spontaneous firing. Firing and resetting of traps in \emph{Utricularia} are associated with water flow and trap volume changes, which mechanism is studied experimentally by Lubomir Adamec. The experimental data suggest that only a direct recording of a physiological activity in trigger hairs of a stimulated \emph{Utricularia} trap can determine whether or not the mechanical stimulus results in an action potential (like in \emph{Dionaea muscipula} and \emph{Aldrovanda vesiculosa}), which could serve as an electrical signal for the trap-door buckling \cite{Adamec2012}. Bartosz J. Plachno \emph{et al.} studied the relationship between trap anatomy and its firing-resetting efficiency in nine Australian \emph{Utricularia} species and found that the trap firing-resetting rate, while conceivably dependent on trap thickness, did not seem to depend on the number of cell layers in the trap wall, as quantitative functional trap characteristics of the 3-4 cell layer thick traps of \emph{U. volubilis} were similar to those of the 2-3 cell layer thick traps of \emph{U. dichotoma} \cite{Plachno2015}.

\section{Catapult-flypaper traps}
\begin{figure}[th]
\centering
\includegraphics[width=0.9\textwidth]{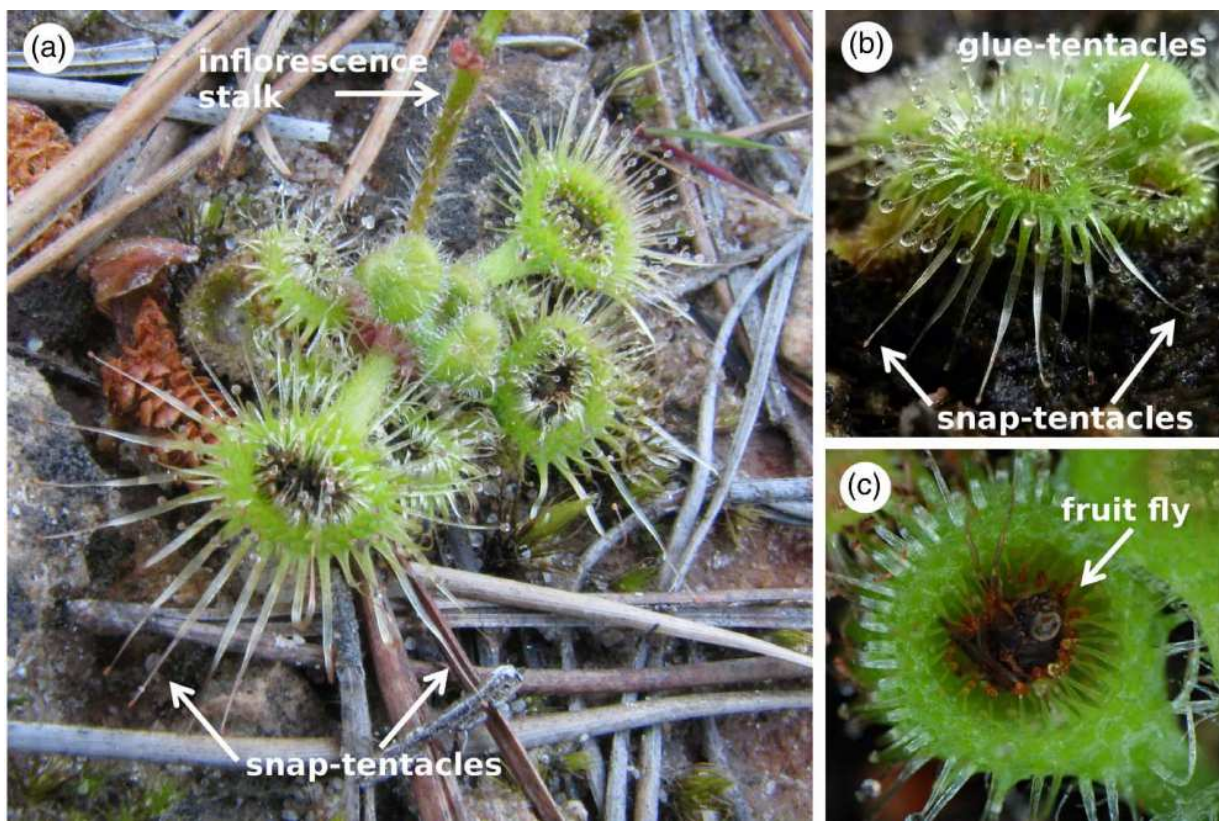}
\caption{Trap leaves of \emph{Drosera glanduligera} (adapted from Ref. \cite{Poppinga2012}). (A) \emph{Drosera glanduligera} in its natural habitat; note the radially long, outward extending, non-sticky snap tentacles and the shorter glue tentacles they snap towards. (B) An example of cultivated \emph{Drosera glanduligera}. (C) \emph{Drosera glanduligera} with a captured fruit fly, stuck within the concave interior of the trap. \label{drosera}}
\end{figure}

Sundews (\emph{Drosera}) are mostly known for capturing prey with sticky traps. However, one recently discovered sundew (\emph{Drosera glanduligera}, \emph{Droseraceae}) demonstrates fast motion for the capture of prey. Upon sensing prey with its touch-sensitive snap tentacle heads, \emph{Drosera glanduligera} catapults prey using outstretched snap-tentacles (as shown in Figure \ref{drosera} (a)) onto adjacent sticky glue-tentacles (Figure \ref{drosera} (b) and (c)) in as little as 75 milliseconds \cite{Hartmeyer2010,Poppinga2012,Hartmeyer2013}. These sticky glue-tentacles then slowly fold up and surround the prey over several minutes to several hours while digesting the prey with secreted digestive enzymes.
The rapid catapulting movement of the snap-tentacle is considered to be primarily due to change in hydraulic pressures within the tentacle. Two possible mechanisms of this hydraulic pressure change have been proposed \cite{Poppinga2012}: rapid transport of water from cells in the adaxial half of the tentacle to cells in the abaxial half, resulting in adaxial contraction and abaxial extension and the fast catapulting movement; or, a sudden loss of turgor pressure in adaxial cells, followed by tentacle bending due to prestress of the abaxial surface to release the elastic energy stored in the epidermal cells on the adaxial side. However, further investigations are needed to test these hypotheses and elucidate the operating mechanisms of the fast catapulting motion of the snap-tentacles. Notably, unlike the snap-through movement of the Venus flytrap, the catapulting movement of the snap-tentacles is irreversible, possibly because non-recoverable fracture of the epidermal cells occurs in the hinge zone of the snap-tentacle upon initiation of movement\cite{Poppinga2012}. Some other flypaper traps, such as \emph{Pinguicula} (\cite{Heslop-Harrison1970,Heslop-Harrison2004}), less mobile in nature, are nevertheless not reviewed in detail in the current article.

\section{Other fast movements in plants}
\begin{figure}[th]
\begin{center}
\includegraphics[width=0.85\textwidth]{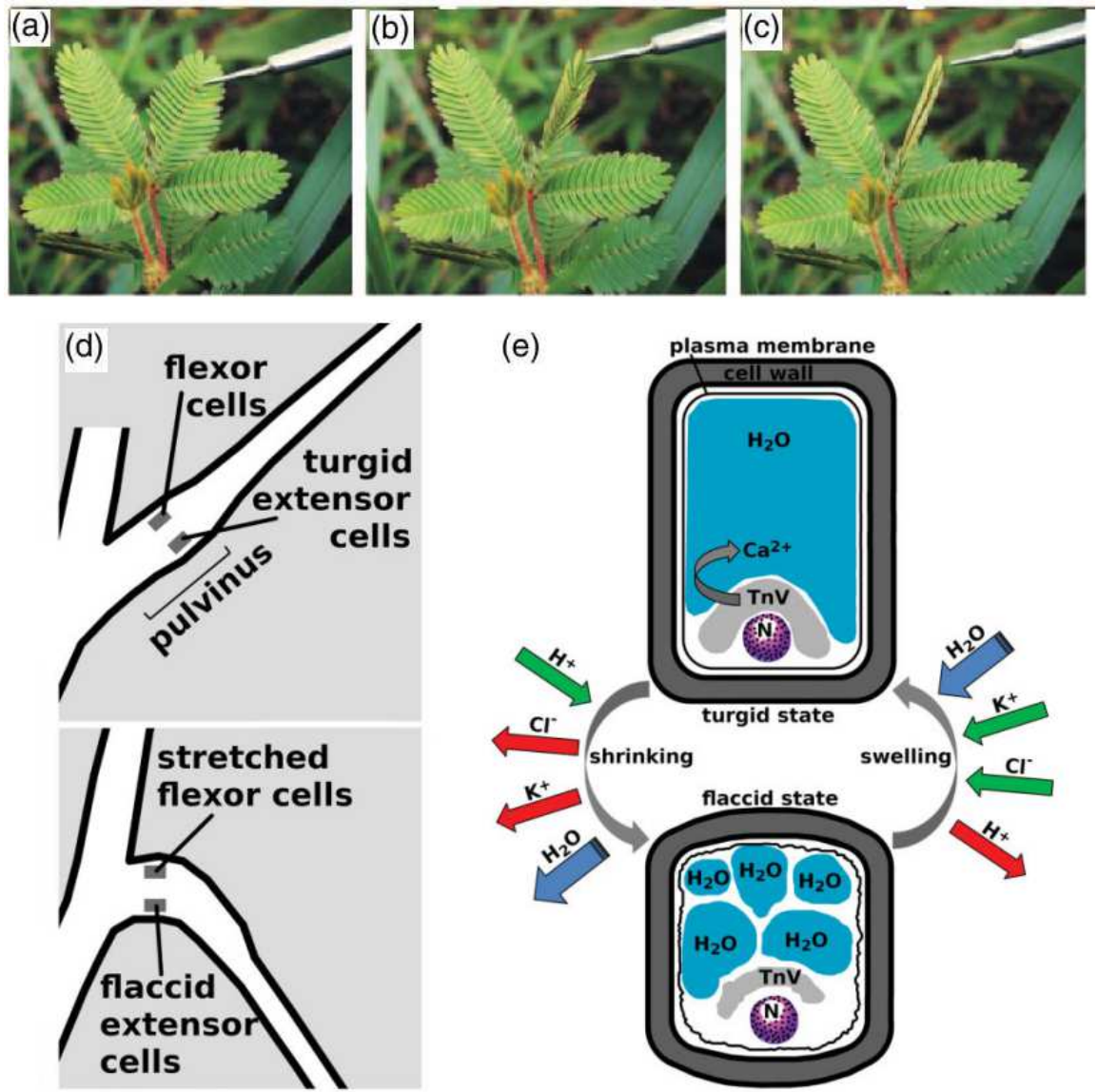}
\end{center}
\caption{Thigmonastic movement of leaflets and schematics of mechanisms in \emph{Mimosa pudica}. (a) Leaflets open; (b) leaflets closing due to touch-induced changes in cell turgor of cells within the pulvinus, a structure located at the base of each leaflet. (c) Leaflets closed. The time-lapse between each photograph is about 1 sec. (a) to (c) are adapted from Ref. \cite{Scorza2011}. (d) Cartoon illustrating the mechanism of leaf movement in \emph{Mimosa pudica} before (top panel) and after (bottom panel) application of stimulus. Note the corresponding states of the flexor and extensor cells in each condition, which implies that the extensor cells are the primary effector of pulvinus bending. (e) Cartoon illustrating turgor changes in an extensor cell and the resulting shape changes of the cell as it changes between turgid (above) and flaccid (below) states; here, N denotes the cell nucleus. The primary ionic movements underlying these turgor changes are shown: when the leaves are mechanically simulated, adenosine triphosphate (ATP) dephosphorylating H$^{+}$ pumps (green) allow for the rapid outward flux of K$^{+}$ and Cl $^{-}$ ions, which in turn trigger outward osmotic movement of water. Tannin-rich vacuoles (TnV) store Ca$^{2+}$ to regulate K$^{+}$ flux. (d) and (e) are adpated from Ref. \cite{Scorza2011} and \cite{Poppinga2013} \label{Mimosa}}
\end{figure}

Besides these carnivorous plants with active traps, there are other plants that can move at a comparably dramatic speed. Since the main focus of this review article is on the fast, nastic movements mostly discovered in carnivorous plants, here we only mention a few examples of other types of fast movements in plants. Interested readers can refer to some excellent review articles \cite{Martone2010,Forterre2013}. As yet another well-known moving plant, \emph{Mimosa pudica} exhibits a rapid, defensive response to external stimuli, such as the closing of its leaves (Figure \ref{Mimosa}(a)-(c)) and bending of its pulvinus, the joint-like thickening of the plant near the bottom of a leaf or leaflet where the plants ``motor cells" reside. \cite{Scorza2011,Poppinga2013} (Figure \ref{Mimosa}(d)). These motor cells, divided into flexor and extensor cells on the ventral and dorsal side of the leaf, respectively, regulate the volume and shape according to their relative turgor pressure. Upon mechanical stimulation of the plant, an action potential is initiated which propagates a signal from the site of the stimulus to the pulvinus. This results in another action potential which initiates the differential transport of K$^{+}$ and Cl$^{-}$, leading to a change in turgor pressure in the cells (Figure \ref{Mimosa}(e)). Consequently, the water potential in the extensor cells increases, resulting in loss of water and shrinkage of the cells, with the opposite occurring in the flexor cells, which swell to a turgid state. As a result, the leaflet folds up, ensnaring the trapped prey within. Calcium ions are shown to play an important role in this nastic movement as well \cite{Toriyama1972,Cote1995,Scorza2011}. More recently, Song \emph{et al.} \cite{Song2014}, using advanced bio-imaging techniques including 2D X-ray micro-imaging and 3D X-ray tomography, non-destructively monitored the pulvinis at a high resolution to analyze the morphological characteristics of the pulvinis in real time. 

In contrast with the plant initiated, hydraulically driven movements of \emph{Mimosa pudica}, some plants use purely mechanical mechanisms to achieve rapid movements. Among such plants, a variety of mechanisms exist to achieve the fast dispersion of small solid objects, such as spores, pollen, or seeds \cite{Edwards2005,Elbaum2007,Martone2010,Forterre2013}. For example, the stamen of the bunchberry dogwood (\emph{Cornus canadensis}) has evolved a trebuchet-like method for ``catapulting" their pollen skywards after the flower opens \cite{Edwards2005}, and the sphagnum moss can shoot its spores out like an airgun using built-up internal pressure \cite{Sundberg2010}. Yet another example of projectile motion is the release of fern spores \cite{King1944,Noblin2012}, where through a cavitation catapult mechanism, generation of bubbles causes rapid cell expansion, inducing dramatic bending of the annulus cells and subsequently converting stored elastic energy into kinetic energy \cite{Noblin2012,Forterre2013}.

Last but not least, a number of plants also exhibit thigmonastic movements in their reproductive parts such as the floral organs and fruits \cite{Jaffe1977, Schlindwein1997,Romero1986, Liu2010, Nicholson2008,Scorza2011,Joyeux2011b}. For example, when touched gently, the stamens of the moss rose, \emph{Portulaca grandiflora} can bend within less than a second and return to the original positions in about five minutes \cite{Jaffe1977}, similar to the thigmonastic movement of \emph{Mimosa} leaves in terms of the response times. Orchids have also evolved remarkable actuating structures and morphologies, such as sexually dimorphic flowers, to induce insects to cross-pollinate for them \cite{Romero1986,Nicholson2008}. Such touch sensitive movements of the reproductive organs represent an active adaption to enhance cross-pollination by transporting the pollen grains to arthropods (such as bees) and birds who serve as pollinators. Interesting questions then naturally arise about how these plants and their pollinators have co-evolved over millions of years. Addressing the molecular and biophysical mechanisms in these thigmonastic movements may help elucidate the co-evolutionary relationships between the species of flowering plants and the pollinators as well as provide information about the evolution paths of these versatile floral morphologies \cite{Scorza2011}.

\section{Engineered actuating structures inspired by fast-moving plants}
Fast-moving plants, as examples of dynamically morphing biological structures, are good model systems in the field of biomimetics, a relatively new discipline that takes inspirations from biology for problems in engineering. For example, the leaves of the \emph{Mimosa} plant, which contract upon certain external stimuli (physical, heat, chemical etc.), have inspired research in structures that can change on demand between flexible and stiff states through changes in hydraulic pressure \cite{Bruhn2014}. In particular, \emph{Dionaea muscipula}, the Venus flytrap, has drawn strong interest from engineers and plant biologists alike. The quick sensing and closing properties of \emph{Dionaea} have inspired a number of imitative designs with potential applications in drug delivery systems, artificial muscles, and shape changing structures \cite{Holmes,Shahinpoor2011,Lee,Chen2012}.

\begin{figure}[th]
\begin{center}
\includegraphics[width=0.9\textwidth]{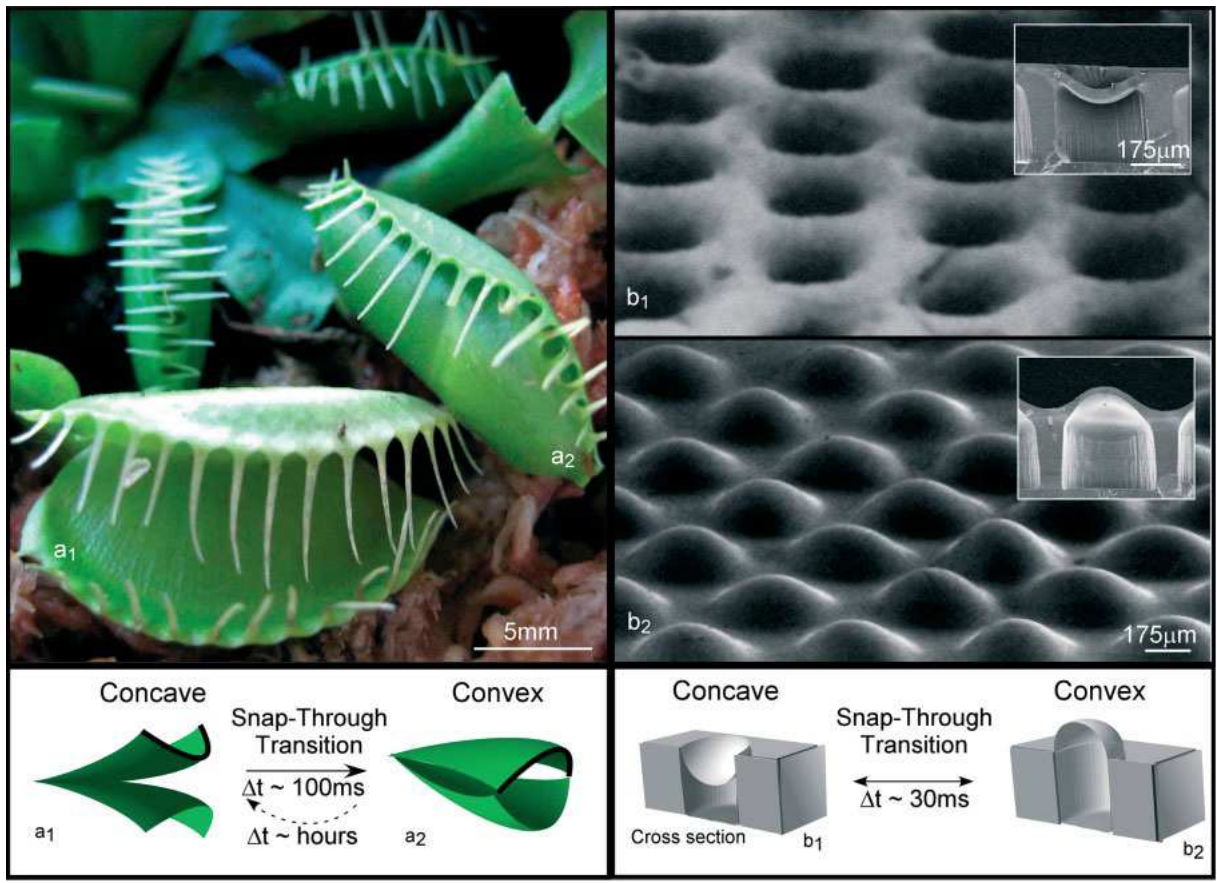}
\end{center}
\caption{Shown is a diagram of ``snap-through" transition of the microlens structure. The surface of a concave microlens (b1) undergoes snap-through transition to a convex shape (b2) in $\sim$30 ms. Figure adapted from Ref. \cite{Holmes}. \label{snapthrough2}}
\end{figure}

Holmes and Crosby \cite{Holmes} used mechanical buckling of plates (in a two-dimensional analog of Euler buckling) and bi-axial compressive loads to fabricate an array of lens-like bistable shells, shown in Figure \ref{snapthrough2}. A layer of biaxially prestretched polydimethylsiloxane (PDMS), which is patterned with periodic arrays of holes, is topped via spin coating by a thin film of uncured PDMS. The two layers of PDMS are then bonded via crosslinking. The micro-lenses formed by this process can possess either a concave or convex shape; a characteristic snap-through movement, similar to that of the Venus flytrap, occurs upon transition between the shapes. The focal point of each micro-lens is either above the structure surface (when the micro-lens is convex) or below the structure surface (when the micro-lens is concave). The researchers demonstrated finely tunable mechanical properties of the micro-lens structure through tuning of the geometric properties of the structure, e.g., lens size and spacing between adjacent lenses.

\begin{figure}[th]
\begin{center}
\includegraphics[width=1.0\textwidth]{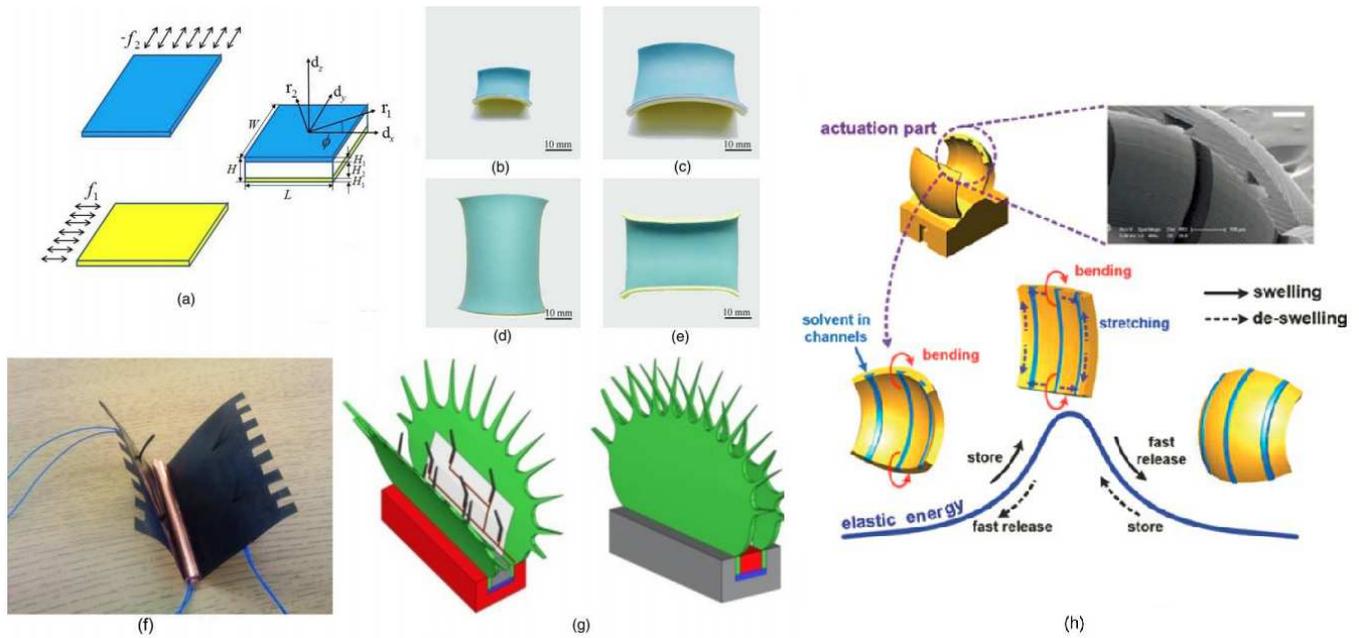}
\end{center}
\caption{(a) Schematics of the bench-top experiments: two pieces of latex rubber sheets (blue and yellow) were pre-stretched
along perpendicular directions and a thicker elastic strip is sandwiched in-between. Upon release, the bonded multilayer
composite will deform into one of the following shapes in (b) to (e): (b) A saddle shape when the composite layer is narrow enough. (c)
A saddle shape when the composite layer is thick enough. (d) A stable, nearly cylindrical shape when the composite layer is wide and thin enough. (e) The other stable shape (bending upwards) for the same sheet as in (d). Figures \ref{Fig6}(a)
to \ref{Fig6}(e) are adapted from Ref. \cite{Chen2012}. (f) The robotic Venus flytrap (VFT) with a pair of lobes (traps), embedded
spine (copper roll), IPMC trigger fingers, and lead wires for sensing and actuation. (g) Schematics of a robotic VFT in
an open and closed configuration (adapted from Ref. \cite{Shahinpoor2011}). (h) Snap-buckling of the doubly curved hydrogel device by swelling of the lobes from solvent in the microfluidic channels. Note the characteristic shape changes and reversal of curvature that occur during actuation of the device. Scale bar indicates 100$\mu m$. Adapted from Ref. \cite{Lee}. \label{Fig6}}
\end{figure}

Chen and coworkers \cite{Chen2012} studied the geometric and mechanical properties of a bistable strip (Figures. \ref{Fig6}(a) to \ref{Fig6}(e)). They proposed a theoretical model for the large deformation of shell structures by the conformation of a strip on the doubly-curved surface of a torus. The theory captures the main feature of the competition between bending and stretching energy of an elastic plate when the Gaussian curvature changes. In addition, they fabricated a physical model of a bistable structure by uniaxially pre-stretching two rubber sheets in perpendicular directions, ``sandwiching" a much thicker rubber sheet between the two pre-stretched sheets and bonding the three sheets together. Using the theoretical model, the researchers identified two dimensionless parameters (related to the mechanical forces and geometry) that control the bistable behavior of the system. In particular, the geometric parameter, $\eta \equiv W \sqrt{\kappa/H}$, also discussed by Armon \emph{et al.} \cite{Armon_2011} in an independent study, is equivalent to $\alpha^{1/4}$ (where $\alpha$ is the geometrical parameter that controls the elastic energy barrier in the Venus flytrap's closure \cite{Forterre}). Their work quantified the conditions for bistability, providing a mathematical framework for the study of bistable and multi-stable structures, with application in artificial muscles, bio-inspired robots, and shape-changing structures. Subsequent finite element simulations and experiments further support the validity of the theory \cite{Guo_2014a,Guo_2014b,Chen_2015}. Based on this mechanical principle, bistable helical ribbons have been designed which can morph from one helical shape to the other upon certain external stimulus \cite{Guo_2014b}.

Hydrogel polymers, which swell and contract in response to a wide range of input stimuli (chemical, temperature, mechanical), can lead to a variety of sensing and actuating systems with a wide range of engineering applications \cite{Geryak}. Inspired by \emph{Dionaea}, Lee \emph{et al.} \cite{Lee} fabricated a jumping microgel device as illustrated in Figure \ref{Fig6}(h), utilizing the sensing-actuating properties of hydrogel to engineer a microgel jumping device that generates a snapping motion within 12 milliseconds. The researchers created a doubly curved shape from hydrogel and employed elastic instability to create the snap-buckling mechanism responsible for the quick actuation of the device. Upon swelling, the microgel legs of the device snap-buckle, resulting in a rapid jump. The power density of this device, at 34.2 mW g$^{-1}$, approaches that of human muscle. Also inspired by \emph{Dionaea}, Gracias and coworkers \cite{Bassik,Yoon} have designed and fabricated smart photo-patterned bilayer hydrogel actuators. These hydrogel actuators exhibit external stimuli dependent swelling behavior, which, employed in a bilayer design, can act as a hinge for a Venus flytrap-like actuator. Santulli and coworkers \cite{Santulli2005}, combining hydrogel polymer with braided fibers, developed and characterized actuators with tunable stiffness and force generation capacity.

In order to understand the behavior of these hydrogel devices, the underlying behaviors and properties of the hydrogel must be characterized and quantified. Stoychev \emph{et al.} \cite{Stoychev2012} explored the folding behavior of rectangular hydrogel bilayers on a substrate, characterizing the effects of both non-homogenous swelling and adhesion of the hydrogel to the substrate.  The capillary forces which drive liquid absorption based hydrogel actuation have been reviewed by Roman and Bico \cite{RomanBico}. Nevertheless, many properties and behaviors of hydrogel structures remain to be explored.

Hydrogel composites allow for the design of complex, dynamically morphing pre-programmed shapes. Erb and coworkers \cite{Erb2013}, inspired by the shape changes in natural seed dispersal units and carnivorous plants, designed hydrogel composites with embedded structural reinforcements which can deform into complex, bioinspired shapes, such as a twisting seed pod. Hydrogels and other polymers may also be utilized with origami design principals, combining the sensing-actuating properties of polymers with the pre-programmed, dynamically tunable properties of origami designs. For instance, Wei \emph{et al.} \cite{Wei2014} designed hybrid hydrogels which can dynamically morph into specific shapes, including helical sheets with reversible chirality, while Ryu and coworkers develped photo-origami designs which, in reaction to certain wavelengths of light, fold into precise 3-D structures from 2-D sheets \cite{Ryu2012}.
Origami designs incorporating stiff plate elements \cite{Kunstler2011} and biomimetic leaf folding patterns \cite{DeFocatiisGuest} might also be used towards the development of novel actuators.

In addition, ionic polymer-metal composites (IPMCs) have garnered much interest from the scientific community for both their biomimetic sensor and actuator properties \cite{Shahinpoor1}. IPMCs exhibit characteristic sensing and mechanical bending in an electrical field, actions which strongly resemble the sensing and bending capabilities of the leaves of the Venus flytrap. IPMCs also have potential medical and industrial applications; current study of IPMCs focuses on further modelling of IPMC actuators and development of application prototypes \cite{Shahinpoor3,Shahinpoor4,Jo2013}. Because of the similarities between the two, studies of the bending and sensing mechanics behind IPMC actuators and Venus flytrap mechanics go hand in hand. As illustrated in Figure \ref{Fig6}(f) and (g), a novel flytrap-inspired robot was fabricated by Shahinpoor using IPMC artificial muscles as distributed sensors and actuators \cite{Shahinpoor2011}. A conductive copper spine served as a ``midrib", which transmitted the signals sensed by the IPMC trigger hair to the solid-state relay system and to activate a small dynamic voltage generator capable of actuating the robotic trap. When certain stimuli were applied to the IPMC trigger hairs, the functioning of the voltage generator can cause the flytrap robot to close. Such biomimetic design of smart systems/structures, integrated with functional sensors and actuators, can be potentially useful in a number of engineering applications \cite{Zheng2013}.

The hydraulic redistribution driven movements of plants have also influenced the development of osmotically actuating, shape changing structures. Sinibaldi \emph{\emph{et al.}} \cite{Sinibaldi2013}, inspired by osmotic-driven plant movements, modelled the dynamic behavior of two implementations of the osmotic actuator concept. Both models were formulated on the basis of two chambers, a reservoir chamber (RC) filled with solute and actuation chamber (AC) filled with solute and solvent, which are separated by a solute-impermeable osmotic membrane (OM). The two models are different in their implementation of the actuator: the first has actuation work stored through the elastic deformation of a spring by a piston which displaces through solvent flux; the second has actuation work stored through the elastic deformation of a non-permeable bulging membrane. These models represent preliminary steps towards understanding towards the development of biomimetic, biorobotic solutions inspired by plant movement.

There are additional examples of morphing structures inspired by the hydraulic redistribution driven plant movement. Inspired by the hydraulic movements of \emph{D. muscipula} and \emph{M. pudica}, Pagitz \emph{et al.} developed a model for morphing structures driven by pressure-actuated cellular structures \cite{Pagitz2012}. The researchers further refined this model by introducing a shell-like structure that can, through its characteristic ``snap-through" morphology, change the sign of its Gaussian curvature \cite{Pagitz2013}. These shape changing structures, driven by multi-layer cellular pressure actuation, have potential applications in aerospace engineering, automobile engineering, architectural design, etc. Moreover, Freeman \cite{Freeman2009, Freeman2012} designed biomimetic flexible morphing membranes, so called ``engineered nastic membranes", for on-demand change of aircraft wing shape during flight. Upon introduction of adenine triphosphate (ATP) to these nastic membranes, proton pumps are activated which move select ions across the membrane into a cylindrical inclusion (analogous to protons being pumped into the intracellular space). The resulting electrochemical gradient activates co-transporters and exchangers located in the nastic membrane, which drive the ions back out of the inclusion and down their electrochemical gradient while concurrently using this ionic flux to move fluid into the inclusion. This inward flow of fluid results in expansion of the cylindrical inclusion until fluid pressures are equalized and hydrostatic equilibrium reached and is responsible for the dynamic morphology of the structure. Detailed analysis of the mechanical, morphological, and electrochemical properties of this bio-inspired actuator in ABAQUS for the surrounding polymer and UFLUID for the driving transport processes revealed that various properties including slow reaction time, difficulty in upwards scaling, and limited displacement prevented this design from application in its original purpose of morphing aircraft wings; however, the design showed promising applications in enhanced osmotic actuation, water purification through protein-driven selective transport and highly targeted drug or vaccine delivery devices, which burst and release their contents upon contact with the proper stimulus in the body. More recent developments to this design by Freeman and Leiland \cite{Freeman2012} incorporated the use of voltage-gated ion channels for the control of activation and a novel dual activation chamber design capable of simultaneous swelling and shrinking, with potential applications in biomorphic structures.

\begin{figure}[th]
\begin{center}
\includegraphics[width=1.0\textwidth]{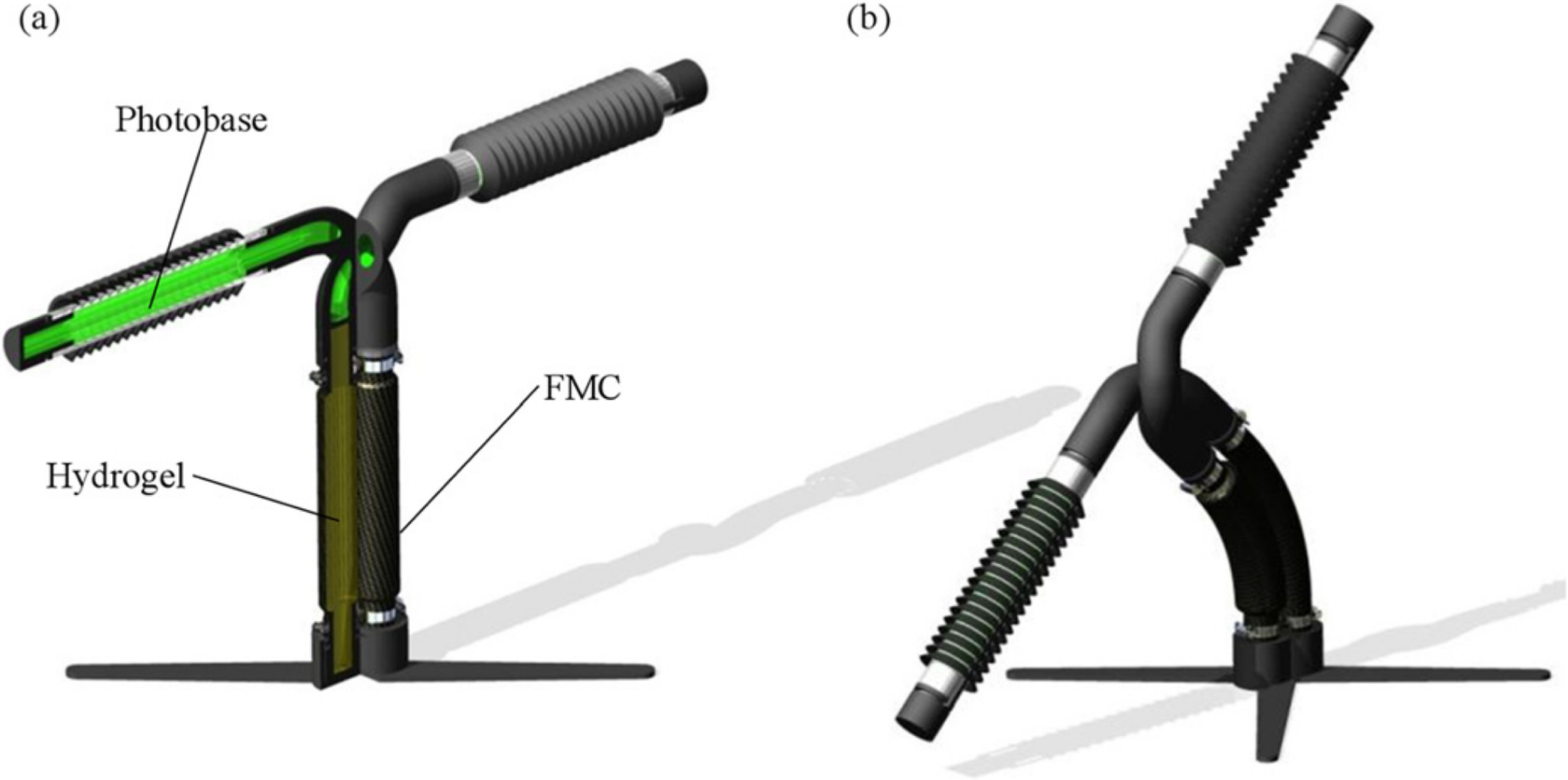}
\end{center}
\caption{(a) Un-actuated device with quarter section removed to reveal internal structure. (b) Full device in actuated state showing orientation towards the light source. Figure adapted from Ref. \cite{Dicker}. \label{suntracker}}
\end{figure}

The mechanical principles demonstrated by moving plants also have potential application in systems engineering. Recently, Lenau and Hesselberg \cite{Lenau2014} studied the self-organization principles demonstrated in stimulated motion plants. They divided the plant motions into three categories (according to the purpose): dispersal of spores, pollen, or seeds (e.g., sphagnum mosses); obtainment of nutrients (e.g., carnivorous plants such as \emph{Dionaea}); and protection of vital organs or obtainment of optimal light and temperature conditions (e.g. \emph{Mimosa}). They outlined the self-organization principles behind sun-tracking plants, i.e., detect sunlight, communicate this signal to an actuator, and perform a motion to optimize internal and external conditions based on the amount of sunlight, and showed that these design principles could be applied to the design of window blinds that open and close based on the amount of sunlight or solar energy devices that track the sun's position through the sky in order to maximize exposure. These designs for sun tracking devices have potential applications in energy conservation on an industrial or consumer scale. Also inspired by the sun-tracking movements of plants, Dicker and coworkers \cite{Dicker} developed a biomimetic analogue of the sun-tracking leaves of the Cornish Mallow (\emph{Lavatera cretica}), which reversibly move the plane of the leaf to be perpendicular to available sunlight, termed diaheliotropic movement. As shown in Figure \ref{suntracker}, the designed biomimetic photo-actuator consists of a reversible photobase as the sensor and pH sensitive hydrogels as the actuator. The photobase 4,4-bis(dimethylamino)triphenlmethane leucohydroxide (malachite green carbinol base) is stored in a pair of artificial leaf veins which limit light exposure via shades oriented 45 degrees to the tube surfaces and changes pH when exposed to UV light. Upon pH change, the pH sensitive hydrogel, an epoxy hydrogel described by Yoshioka and Calvert \cite{YoshiokaCalvert}, reversibly swells within a pair of novel flexible matrix composite (FMC) tubes in a twisted double T joint configuration. Due to the twisted double T joint configuration of the tubes and limited light exposure of the artificial leaf veins, the biomimetic photo-actuator continues to bend until the light exposure of the photobase is maximized, analogous to the sun-tracking movements of plants. This biomimetic sun-tracking device acts as a proof of concept for chemical sensing, control and actuation with potential applications in solar tracking for solar power generation.

\section{Conclusion}
Plants with fast thigmonastic movements have evolved the capability to quickly and dramatically alter their shapes in response to certain external stimuli. Although these mechanically-induced active movements may appear to serve different purposes, from a protective mechanism (e.g., the folding of $Mimosa$ leaflets) to a hunting function (for example, the rapid closure of the traps of $\emph{Dionaea}$, $\emph{Utricularia}$, and $\emph{Aldrovanda}$), some common ground exists. For instance, recent investigations have started to show that most thigmonastic movements result from differential changes of turgor pressure in the tissues, and the opening of certain key ion channels are often involved \cite{Scorza2011}. While some biophysical and biochemical mechanisms behind the fast motions of plants have been extensively studied, many questions remain unanswered. For example, how the action potentials are generated through the touch receptors in $Dionaea$ remains to be investigated. Moreover, the source and role of action potentials in fast-moving plants, especially in carnivores other than $Dionaea$, are not completely understood. The coupling between these action potentials and molecular-level activities (which drive macroscopic movement) also remains not well understood. In $\emph{Dionaea}$, recent studies suggest that it is the speed of ion transport, instead of water transport, that constrains the speed of osmotic flows; therefore, alternative mechanisms of perturbing the water potential in cells should be searched for \cite{Forterre2013}. Last but not least, the roles of actin cytoskeleton, microtublues, and auxin during these active movements require elucidation. 

Understanding the molecular, biomechanical, biochemical, and electrophysical mechanisms among the diverse types of movements in plants may be the key to reveal the universal mechanisms governing plants' movements and how they evolve as a result of adaption to the ever-changing living environments \cite{Haswell2013}. However, it remains a grand challenge how to develop more realistic, multi-scale biomechanical models that incorporate various factors at the molecular, cellular, tissue, and organism level. The difficulty in bridging the physical and biological aspects of the problem also arises from the fact that there is a lack of effective non-invasive experimental techniques for measuring the stresses inside cells and monitoring the transport of fluids (not to mention ions) inside the plant cells during these rapid movements \cite{Forterre2013}. The recent \emph{in vivo} measurement of pressures and poroelastic properties in cells of $Dionaea$ \cite{ColombaniForterre} and the measurement of mechanical stresses in developing embryos \cite{Campas2014} seem to have paved way for further breakthroughs in investigating the underlying mechanisms in fast-moving plants. In addition, the comparisons between the behaviors and operating mechanisms in the moving plants, $Mimosa$, $Dionaea$, $Aldrovanda$, $Utricularia$, and $Drosera$ also bring forth interesting questions from the evolutionary biology and comparative biomechanics perspectives that remain to be addressed. The study of fast-moving plants will not only foster integration of physics, chemistry, biomechanics, and molecular biology to address intriguing mechanisms involved in plants' fast motions, but also inspire various kinds of biomimetic designs for engineering applications.

\section*{Author Contributions}
Q.G., E.D., X. H., and Z.C. generated the idea for the manuscript, performed literature searches, helped to draft and edit it. S.X. and E.C. also performed literature searches, helped to draft and edit it.

\section*{Acknowledgments}
The authors thank J. Hanlin, I. Trase, J. Ballard, K. Tan, H. Zheng, and the anonymous reviewers for helpful comments.

\section*{Funding statement}
This work is in part supported by the National Science Foundation of China Grant No.11102040, Projects of International Cooperation and Exchanges NSFC (Grant No.11201001044), and Fujian University of Technology Research Fund （Grant No.GY-Z15006）. Z.C. acknowledges support from the Society in Science - Branco Weiss fellowship, administered by ETH Z$\ddot{u}$rich.

\section*{Conflict of interests} The authors have no financial or personal relationships with other people or organizations that might inappropriately influence or bias the work.

%
%

%




\begin{thebibliography}{100}

	\bibitem{Darwin1880} Darwin C. 1880. The Power of Movement in Plants. London: William Clowes.

	\bibitem{Goriely1998} Goriely A., and Tabor M. 1998 Spontaneous helix-hand reversal and tendril perversion in climbing plants. \emph{Phys. Rev. Lett.} \textbf{80}, 1564-1568. (doi: 10.1103/PhysRevLett.80.1564)

	\bibitem{Gerbode2012} Gerbode SJ, Puzey JR, McCormick AG, Mahadevan L. 2012 \emph{Science} \textbf {337}, 1087-1091. (doi: 10.1126/science.1223304)

	\bibitem{Wang2013} Wang JS, Wang G., Feng XQ, Kitamura T, Kang YL, Yu SW, Qin QH. 2013 Hierarchical chirality transfer in the growth of towel gourd tendrils \textbf{Sci. Rep.} \textbf{3}, 3102 (2013). (doi: 10.1038/srep03102)

	\bibitem{Silverberg2012} Silverberg JL, Noar RD, Packer MS, Harrison MJ, Henley CL, Cohen I., Gerbode SJ. 2012 3D imaging and mechanical modeling of helical buckling in \emph{Medicago truncatula} plant roots. \emph{Proc. Natl. Acad. Sci.} \textbf{109}, 16794-16799. (doi: 10.1073/pnas.1209287109)

	\bibitem{Darwin} Darwin C. 1875. Insectivorous Plants. London: Murry.

	\bibitem{Ueda2006} Ueda M and Nakamura Y. 2006 Metabolites involved in plant movement and `memory': nyctinasty of legumes and trap movement in the Venus flytrap. \emph{Nat. Prod. Rep.} \textbf{23}, 548-557. (doi:10.1039/b515708k)

	\bibitem{Fratl2009} Fratl P, Barth FG. 2009 Biomaterial systems for mechanosensing and actuation. \emph{Nature} \textbf{462}, 442-448. (doi:10.1038/nature08603)

	\bibitem{Ellison2009} Ellison AM, Gotelli NJ. 2009. Energetics and the evolution of carnivorous plants-Darwin's `most wonderful plants in the world'. \emph{J. Exp. Bot.} \textbf{60}, 19-42. (doi: 10.1093/jxb/ern179)

	\bibitem{Martone2010} Martone PT, Boller M, Burgert I, Dumais J, Edwards J, Mach K, Rowe N, Rueggeberg M, Seidel R, Speck T. 2010 Mechanics without muscle: biomechanical inspiration from the plant world. \emph{Integr. Comp. Biol.} \textbf{50}, 888. (doi: 10.1093/icb/icq122)

	\bibitem{Joyeux2011b} Joyeux M. 2012 At the conjunction of biology, chemistry and physics: the fast movements of \emph{Dionaea}, \emph{Aldrovanda}, \emph{Utricularia} and \emph{Stylidium}. \emph{Frontiers in Life Science} \emph{5}, 71-79 (doi: 10.1080/21553769.2012.708907)

	\bibitem{Dumais2012} Dumais J, Forterre Y. 2012 ``Vegetable dynamicks": the role of water in plant movements. \emph{Annual Review of Fluid Mechanics} \textbf{44}, 453-478. (DOI: 10.1146/annurev-fluid-120710-101200)

	\bibitem{Zheng2013} Zheng H, Liu Y, Chen Z. 2013 Fast Motion of Plants: from Biomechanics to Biomimetics. \emph{J. P. Res.} \textbf{1}, 40-50. (10.14304/SURYA.JPR.V1N2.5)

	\bibitem{Forterre2013} Forterre Y 2013 Slow, fast and furious: understanding the physics of plant movements. \emph{J. Exp. Bot.} \textbf{64}, 4745-4760. (doi: 10.1093/jxb/ert230)

	\bibitem{Poppinga2013} Poppinga S, Masselter T, Spec T. 2013 Faster than their prey: New insights into the rapid movements of active carnivorous plants traps. \emph{BioEssays} \textbf{35}, 649-657. (doi: 10.1002/bies.201200175)

	\bibitem{Moulia2013} Moulia B. 2013 Plant biomechanics and mechanobiology are convergent paths to flourishing interdisciplinary research. \emph{J. Exp. Bot.} \textbf{64}, 4617-4633. (doi:10.1093/jxb/ert320)

	\bibitem{Discher2005} Discher DE, Janmey P, Wang YL. 2005 Tissue cells feel and respond to the stiffness of their substrate. \emph{Science} textbf{310}, 1139-1143. (doi: 10.1126/science.1116995b)

	\bibitem{Savin2011} Savin T, Kurpios NA, Shyer AE, Florescu P, Liang H, Mahadevan L, Tabin CJ. 2011 On the growth and form of the gut. \emph{Nature} \textbf{476} 57-62. (doi: 10.1038/nature10277)

	\bibitem{Wyczalkowski2012} Myczalkowski MA, Chen Z, Filas BA, Varner VD, Taber LA. 2012 Computational models for mechanics of morphogenesis. \emph{Birth Defects Research Part C: Embryo Today: Reviews}. \textbf{96}, 2.

	\bibitem{Kuhl2013} Kuhl E. Growing matter: A review of growth in living systems. \emph{J. Mech. Behav. Biomed. Mater.} \textbf{29}, 529-543. (doi:10.1016/j.jmbbm.2013.10.009)

	\bibitem{Gleghorn2013} Gleghorn J, Manivannan S, Nelson CM. 2013 Quantitative approaches to uncover physical mechanisms of tissue morphogenesis. \emph{Curr. Opin. Biotech.} \textbf{24}, 954-961. (doi:10.1016/j.copbio.2013.04.006)

	\bibitem{Chihan2013} Prentice-Motta HV, Chang CH, Mahadevan L, Mitchisona TJ, Irimiag D, Shah JV. 2013 Biased migration of confined neutrophil-like cells in asymmetric hydraulic environments. \emph{Proc. Natl. Acad. Sci.} \textbf{110}, 21006-21011. (doi: 10.1073/pnas.1317441110)

	\bibitem{Braam2005} Braam J. In touch: plant responses to mechanical stimuli. \emph{New Phytologist} \textbf{165}, 373-389. (doi: 10.1111/j.1469-8137.2004.01263.x)

	\bibitem{Burgert2009} Burgert I, Fratzl P. 2009 Actuation systems in plants as prototypes for bioinspired devices. \emph{Philos. T. R. Soc. A} \textbf{367}, 1541-1557. (doi: 10.1098/rsta.2009.0003)

	\bibitem{Scorza2011} Scorza LCT, Dornelas MC. 2011 Plants on the move. Toward common
mechanisms governing mechanically-induced plant movements. \emph{Plant Signal. Behav.} \textbf{6}, 1979-86. (doi: 10.4161/psb.6.12.18192)

	\bibitem{Hill1981} Hill BS, Findlay GP. 1981 The power of movement in plants: the role of osmotic machines. \emph{Q Rev Biophys.} \textbf{14} 173-222.

	\bibitem{Elbaum2007} Elbaum R, Zaltzman L, Burgert I, Fratzl P. 2007. The role of wheat awns in the seed dispersal unit. \emph{Science} \textbf{316}, 884-886. (doi: 10.1098/rsta.2009.0003)

	\bibitem{Armon_2011} Armon S, Efrati E, Kupferman R, Sharon E. 2011 Geometry and mechanics in the opening of chiral seed pods \textbf{Science} \textbf{333}, 1726-30. (doi: 10.1126/science.1203874)

	\bibitem{Guo2013} Guo Q, Zheng H, Chen W, Chen Z, Finite Element Simulations on Mechanical Self-assembly of Biomimetic Helical Structures. \emph{J. Mech. Med. Biol.} \textbf{13}, 1340018 (2013). (doi: 10.1142/S0219519413400186)

	\bibitem{Volkov2008} Volkov AG, Adesina T, Markin VS, Jovanov E. 2008. Kinetics and mechanism of \emph{Dionaea muscipula} trap closing. \emph{Plant Physiol.} \textbf{146}, 694-702. (doi: 10.1104/pp.107.108241)

	\bibitem{Brown1910} Brown WH, Sharp LW. 1910 The Closing Response in \emph{Dionaea}. \emph{Botanical Gazette} \textbf{Vol. 49, No. 4}, 290-302.

	\bibitem{Volkov2007} Volkov AG, Adesina T, Jovanov E. 2007. Closing of Venus flytrap by electrical stimulation of motor cells. \emph{Plant Signa. Behav.}, \textbf{2}, 139-145.

	\bibitem{Volkov2011} Volkov AG, Pinnock MR, Lowe DC, Gay MS, Markin VS. 2011 Complete hunting cycle of \emph{Dionaea muscipula} Consecutive steps and their electrical properties. \emph {J Plant Physiol.} \emph{168}, 109-20. (doi: 10.1016/j.jplph.2010.06.007)

	\bibitem{DiPalma} DiPalma JR, McMichael R, DiPalma M. 1966. Touch receptor of Venus flytrap, \emph{Dionaea muscipula}. \emph{Science} \textbf{152}, 539-540. (doi:10.1126/science.152.3721.539)

	\bibitem{Forterre} Forterre Y, Skotheim JM, Dumais J, Mahadevan L. 2005. How the Venus flytrap snaps. \emph{Nature} \textbf{433}, 421-425. (doi: 10.1038/nature03185)

	\bibitem{Yang2010} Ruoting Y, Lenaghan SC, Zhang M, Lijin Xia L. 2010 A Mathematical Model on the Closing and Opening Mechanism for Venus Flytrap, \emph{Plant Signal. Behav.}, \textbf{5}, 968-978. (doi: 10.4161/psb.5.8.12136)

	\bibitem{Brown} Brown WH. 1916. The mechanism of movement and the duration of the effect of stimulation in the leaves of \emph{Dionaea}. \emph{Am. J. Bot.} \textbf{3}, 68-90.( Stable URL: http://www.jstor.org.libproxy.wustl.edu/stable/2435207)

	\bibitem{ColombaniForterre} Colombani M, Forterre Y. 2011 Biomechanics of rapid movements in plants: poroelastic measurements at the cell scale. \emph{Computer Methods in Biomechanics and Biomedical Engineering} \textbf{14} S1, 115-117.

	\bibitem{Markin} Markin VS, Volkov AG, Jovanov E. 2008. Active movements in plants: mechanism of trap closure by \emph{Dionaea muscipula Ellis}. \emph{Plant Signa. Behav.}, \textbf{3}, 778-783. (doi: 10.4161/psb.3.10.6041)

	\bibitem{Joyeux2013} Joyeux M. 2013. Elastic models of the fast traps of carnivorous \emph{Dionaea} and \emph{Aldrovanda}. \emph{Phys. Rev. E} \textbf{88}, 034701. (doi: 10.1103/PhysRevE.88.034701)

	\bibitem{Joyeux2011} Joyeux M, Vincent O, Marmottant P. 2011. Mechanical model of the ultrafast underwater trap of \emph{Utricularia}. \emph{Phys. Rev. E} \textbf{83}, 021911. (doi:10.1103/PhysRevE.83.021911)

	\bibitem{Pandolfi2014} Pandolfi C, Masi E, Voigt B, Mugnai S, Volkmann D, Mancuso S. 2014 Gravity Affects the Closure of Traps in \emph{Dionaea muscipula}. \emph{BioMed Research International} \textbf{2014}, 964203. (doi: 10.1155/2014/964203)

	\bibitem{Volkov2009} Volkov AG, Carrell H, Baldwin A, Markin VS. 2009. Electrical memory in Venus flytrap. \emph{Bioelectrochemistry} \textbf{75},142-147.(doi: 10.1016/j.bioelechem.2009.03.005)

	\bibitem{Eisen} Eisen D, Janssen D, Chen X, Choa F, Kostov D, Fan J. 2013 Closing a Venus Flytrap with electrical and mid-IR photon stimulations. \emph{Proc. of SPIE} 85655I(1-10). (doi:10.1117/12.2005351)

	\bibitem{Volkov2014} Volkov AG, Forde-Tuckett V, Volkova MI. 2014 Morphing Structures of the \emph{Dionaea muscipula Ellis} during the trap opening and closing. \emph{Plant Signa. Behav.}, \textbf{9}, e27793. (doi: 10.4161/psb.27793)

	\bibitem{Perez} Escalante-P\'erez M, Krol E, Stange A, Geiger D, Al-Rasheid KA, Hause B, Neher E, Hedrich R. 2011. A special pair of phytohormones controls excitability, slow closure, and external stomach formation in the Venus flytrap. \emph{Proc. Natl. Acad. Sci.} \textbf{108}, 15492-15497. (doi: 10.1073/pnas.1112535108)

	\bibitem{Ueda2010} Ueda M, Tokunaga T, Okada M, Nakamura Y, Takada N, Suzuki R, Kondo K. 2010. Trap-closing Chemical Factors of the Venus Flytrap (\emph{Dionaea muscipulla Ellis}). \emph{Chembiochem}, \textbf{11}, 2378-2383. (doi: 10.1002/cbic.201000392)

	\bibitem{Volkov2009b} Volkov AG, Carrell H, Markin VS. 2009 Molecular electronics of the \emph{Dionaea muscipula} trap. \emph{Plant Physiol.} \emph{149}, 1661-7. (doi: 10.1104/pp.108.134536)

	\bibitem{Rainier} R Stahlberg. 2006. Historical Overview on Plant Neurobiology. \emph{Plant Signlaing \& Behavior.} \emph{1:1}, 6-8.

	\bibitem{Poppinga2011} Poppinga S, Joyeux M. 2011 Different mechanics of snap-trapping in the two closely related carnivorous plants \emph{Dionaea muscipula} and \emph{Aldrovanda vesiculosa}. \emph{Phys. Rev. E} \textbf{84}, 041928. (doi: 10.1103/PhysRevE.84.041928)

	\bibitem{Skotheim2005} Skotheim JM, Mahadevan L. 2005. Physical limits and design principles for plant and fungal movements. \emph{Science.} \textbf{308}, 1308-1310. (doi: 10.1126/science.1107976)

	\bibitem{Vincent1} Vincent O, Weibkopf C, Poppinga S, Masselter T, Speck T, Joyeux M, Quilliet C, Marmottant P. 2011. Ultra-fast underwater suction traps. \emph{P. Roy. Soc. Lond. B Bio.} \textbf{278}, 2909-2914.

    \bibitem{Sydenham1985} Sydenham PH, Findlay GP. 1973 The rapid movement ofthe bladder of \emph{Utricularia}
sp. \emph{Aust. J. Biol. Sci.} \textbf{26}, 1115–1126.

	\bibitem{Singh2011} Singh AK, Prabhakar S, Sane SP. 2011 The biomechanics of fast prey capture in aquatic bladderworts. \emph{Biology Letters} \textbf{7}, 547-550. (doi:  10.1098/rsbl.2011.0057)


	\bibitem{Vincent2} Vincent O, Roditchev I, Marmottant P. 2011. Spontaneous firings of carnivorous aquatic \emph{Utricularia} traps: temporal patterns and mechanical oscillations. \emph{PloS one} \textbf{6}, e20205. (doi:10.1371/journal.pone.0020205)

	\bibitem{Adamec2012} Adamec L. 2012 Firing and resetting characteristics of carnivorous \emph{Utricularia} reflexa traps: physiological or only physical regulation of trap triggering. \emph{Phyton}, \textbf{52}, 281-290.

	\bibitem{Plachno2015} Plachno BJ, Adamec L, Kaminska I. 2015 Relationship between trap anatomy and function in Australian carnivorous bladderworts (\emph{Utricularia}) of the subgenus \emph{Polypompholyx}. \emph{Aquatic Botany}, \textbf{120}, 290-296.

	\bibitem{Hartmeyer2010} Hartmeyer I, Hartmeyer SRH. 2010 Snap-tentacles and runway lights. \emph{Carniv Pl Newslett} \textbf{39}, 101-113.

	\bibitem{Poppinga2012} Poppinga S, Hartmeyer SRH, Seidel R, Masselter T. \emph{et al.} 2012 Catapulting tentacles in a sticky carnivorous plant. \emph{PLoS ONE} \textbf{7}, e45735. (10.1371/journal.pone.0045735)

	\bibitem{Hartmeyer2013} Hartmeyer I, Hartmeyer SRH, Masselter T, Seidel R. \emph{et al.} 2013 Catapults into a deadly trap: the unique prey-capture mechanism of \emph{Drosera glanduligera}. \emph{Carniv Pl Newslett.} \textbf{42}, 4-14.

    \bibitem{Heslop-Harrison1970} Heslop-Harrison Y. 1970 Scanning electron microscopy of fresh leaves of Pinguicula. \emph{Science}, \textbf{167}, 172–173.
    \bibitem{Heslop-Harrison2004} Heslop-Harrison Y. 2004 Biological ﬂora of the British Isles. \emph{Pinguicula} L. \emph{J. Ecol.}, \textbf{2004}, \textbf{92}, 1071-1118.
	\bibitem{Toriyama1972} Toriyama H, Jaffe MJ. 1972 Migration of calcium and its role in the regulation of seismonasty in the motor cell of \emph{Mimosa pudica L.} \emph{Plant Physiol.} \textbf{49}, 72-81. (DOI:10.1104/pp.49.1.72)

	\bibitem{Cote1995} Cote GG. 1995 Signal-transduction in leaf movement. \emph{Plant Physiol.} \textbf{109}, 729-34. (PMID:12228627)

	\bibitem{Song2014} Song K, Yeom E, Lee SJ. 2014 Real-time imaging of pulvinus bending in \emph{Mimosa pudica} \emph{Sci. Rep.} \textbf{4}, 6466. (doi:10.1038/srep06466)

	\bibitem{Edwards2005} Edwards J, Whitaker D, Klionsky S, Laskowski MJ. 2005 A record-breaking pollen catapult. \emph{Nature} \textbf{435}, 164. (doi:10.1038/435164a)

	\bibitem{Sundberg2010} Sundberg S. 2010 Size matters for violent discharge height and settling speed of Sphagnum spores: Important attributes for dispersal potential. \emph{Ann Bot.} \textbf{105}, 291-300. (doi: 10.1093/aob/mcp288)

	\bibitem{King1944} King AL. 1944 The spore discharge mechanism of common ferns. \emph{Proc. Natl. Acad. Sci.} \textbf{30}, 155-161. (PMCID: PMC1078688)

	\bibitem{Noblin2012} Noblin X, Rojas NO, Westbrook J, Llorens C, Argentina M, Dumais J. 2012 The fern sporangium: a unique catapult. \emph{Science} \textbf{335}, 1322. (doi:10.1126/science.1215985)

	\bibitem{Jaffe1977} Jaffe MJ, Gibson C, Biro R. 1977 Physiological Studies of Mechanically Stimulated Motor Responses of Flower Parts. 1. Characterization of Thigmotropic Stamens of Portulaca-Grandiflora Hook. \emph{Bot. Gaz.} \textbf{138}, 438-47. (doi:10.1086/336946)

	\bibitem{Schlindwein1997} Schlindwein C, Wittmann D. 1997 Stamen movements in flowers of Opuntia (Cactaceae) favour oligolectic pollinators. \emph{Plant Syst. Evol.} \textbf{204}, 179-93. (doi:10.1007/BF00989204)

	\bibitem{Romero1986} Romero GA, Nelson CE. 1986 Sexual dimorphism in Catasetum orchids forcible pollen emplacement and male flower competition. \emph{Science} \textbf{232}, 1538-40. (doi:10.1126/science.232.4757.1538)

	\bibitem{Liu2010} Liu ZJ, Chen LJ, Liu KW, Li LQ, Rao WH. 2010 A floral organ moving like a caterpillar for pollinating. \emph{J. Syst. Evol.} 48:102-8. (doi:10.1111/j.1759-6831.2009.00065.x)

	\bibitem{Nicholson2008} Nicholson CC, Bales JW, Palmer-Fortune JE, Nicholson RG. 2008 Darwins bee-trap: the kinetics of Catasetum, a new world orchid. \emph{Plant Signal. Behav.} \textbf{3}, 19-23. (doi:10.4161/psb.3.1.4980)

	\bibitem{Bruhn2014} Bruhn BR, Schroeder TBH, Li S., Billeh YN, Wang KW, Mayer M. 2014 Osmosis-Based Pressure Generation: Dynamics and Application. \emph{PLOS One} \textbf{9}, e91350. (doi:10.1371/journal.pone.0091350)

	\bibitem{Holmes} Holmes DP, Ursiny M, Crosby AJ. 2007. Crumpled surface structures. emph{Soft Matter} \textbf{4}, 82-85. (doi: 10.1039/b712324h)

	\bibitem{Shahinpoor2011} Shahinpoor M. 2011. Biomimetic robotic Venus flytrap (\emph{Dionaea muscipula Ellis}) made with ionic polymer metal composites. \emph{Bioinspir. Biomim.} \textbf{6}, 046004(1-11). (doi:10.1088/1748-3182/6/4/046004)

	\bibitem{Lee} Lee H, Xia C, Fang NX. 2010. First jump of microgel; actuation speed enhancement by elastic instability. \emph{Soft Matter} \textbf{6}, 4342-4345. (doi: 10.1039/c9sm00092b)

	\bibitem{Chen2012} Chen Z, Guo Q, Majidi C, Chen W, Srolovitz DJ, Haataja MP. 2012. Nonlinear Geometric Effects in Mechanical Bistable Morphing Structures. Phys. Rev. Lett. \emph{109}, 114302. (doi: 10.1103/PhysRevLett.109.114302)

	\bibitem{Guo_2014a} Guo Q, Zheng H, Chen W, Chen Z. 2014. Modeling bistable behaviors in morphing structures through finite element simulations. \emph{Biomed. Mater. Eng.} \textbf{24}, 557-562. (doi:10.3233/BME-130842)

	\bibitem{Guo_2014b} Guo Q, Mehta AK, Grover MA, Chen W, Lynn DG, Chen Z. 2014. Shape selection and multi-stability in helical ribbons. \emph{Appl. Phys. Lett} \textbf{104}, 211901. (doi:10.1063/1.4878941)

	\bibitem{Chen_2015} Chen Z. Shape Transition and Multi-stability of Helical Ribbons: a Finite Element Method Study. \emph{Archive of Applied Mechanics}, in press (doi:10.1007/s00419-014-0967-2).

	\bibitem{Geryak} Geryak R, Tsukruk VV. 2014 Reconfigurable and actuating structures from soft materials. \emph{Soft Matter} \textbf{10}, 1246-63. (doi: 10.1039/c3sm51768c)

	\bibitem{Bassik} Bassik N, Abebec BT, Laflina KE, Gracias DH. 2014 Photolithographically patterned smart hydrogel based bilayer actuators. \emph{Polymer Communication} \textbf{51}, 6093-6098. (doi:10.1016/j.polymer.2010.10.035)

	\bibitem{Yoon} Yoon C, Xiao R., Park J., Cha J., Nguyen TD, Gracias DH. 2014 Functional stimuli responsive hydrogel devices by self-folding. \emph{Smart Materials and Structures} \textbf{23}, 094008. (doi:10.1088/0964-1726/23/9/094008)

	\bibitem{Santulli2005} Santulli C, Patel SI, Jeronimidis G, Davis FJ, Mitchell GR. 2005 Development of smart variable stiffness actuators using polymer hydrogels. \emph{Smart Mater. Struct.} \textbf{14}. 434-440. (doi:10.1088/0964-1726/14/2/018)

	\bibitem{Stoychev2012} Stoychev G, Zakharchenko S, Turcaud S, Dunlop JWC, Ionov L. 2012 Shape-Programmed Folding of Stimuli-Responsive Polymer Bilayers. \emph{ACS Nano.} \textbf{6}, 3925-3924. (doi:10.1021/nn300079f)

	\bibitem{RomanBico} Roman B and Bico J. 2010 Elasto-capillarity: deforming an elastic structure with a liquid droplet. \emph{ J. Phys.: Condens. Matter.} \textbf{22}, 493101. (doi:10.1088/0953-8984/22/49/493101)

	\bibitem{Erb2013} Erb RM, Sander JS, Grisch R, Studart AR. 2013 Self-shaping composites with programmable bioinspired microstructures. \emph{Nature Communications.} \textbf{4}, 1712. (doi:10.1038/ncomms2666)

	\bibitem{Wei2014} Wei Z, Jia Z, Athas J, Wang C, Raghavan SR, Li T, Nie Z. 2014 Hybrid hydrogel sheets that undergo pre-programmed shape transformations. \emph{Soft Matter.}. \textbf{10}, 8157-8162. (doi:10.1039/C4SM01299B)

	\bibitem{Ryu2012} Ryu J, D'Amato M, Cui X, Long KN, Qi HJ, Dunn ML. 2012 Photo-origami: Bending and folding polymers with light. \emph{Appl. Phys. Lett.} \textbf{100}, 161908. (doi:10.1063/1.3700719)

	\bibitem{Kunstler2011} K\"unstler A and Trautz M. 2011 Wandelbare Faltungen aus biegesteifen Faltelementen. \emph{Bautechnik.} \textbf{88}, 86-93. (doi: 10.1002/bate.201110008)

	\bibitem{DeFocatiisGuest} De Focatiis DSA, Guest SD. 2002 Deployable membranes designed from folding tree leaves. \emph{Phil. Trans. A.} 16214680. (doi:10.1098/rsta.2001.0928)

	\bibitem{Shahinpoor1} Shahinpoor M, Kim KJ. 2001 Ionic polymer-metal composites: I. Fundamentals. \emph{Smart Mat. Struct.} \textbf{10}, 819. (doi:10.1088/0964-1726/10/4/327)

	\bibitem{Shahinpoor3} Shahinpoor M, Kim KJ. 2004. Ionic polymer-metal composites: III. Modeling and simulation as biomimetic sensors, actuators, transducers, and artificial muscles. \emph{Smart mater. Struct.} \textbf{13}, 1362. (doi:10.1088/0964-1726/13/6/009)

	\bibitem{Shahinpoor4} Shahinpoor M, Kim KJ. 2005. Ionic polymer-metal composites: IV. Industrial and medical applications. \emph{Smart mater. Struct.} \textbf{14}, 197. (doi:10.1088/0964-1726/14/1/020)

	\bibitem{Jo2013} Jo C, Pugal D, Oh I, Kim KJ, Asaka K. 2013. Recent advances in ionic polymer metal composite actuators and their modeling and applications. \emph{Prog. Polym. Sci.} \textbf{38}, 1037-1066. ( doi:10.1016/j.progpolymsci.2013.04.003)

	\bibitem{Sinibaldi2013} Sinibaldi E., Puleo GL, Mattioli F., Mattoli V., Di Michele F., Beccai L., Tramaere F., Mancuso S., Mazzolai B. 2012 Osmotic actuation modelling for innovative biorobotic solutions inspired by the plant kingdom. \emph{Bioinspr. Biomim} \textbf{8} 2025002. (doi:10.1088/1748-3182/8/2/025002)

	\bibitem{Pagitz2012} Pagitz M, Lamacchia E, Hol J. 2012. Pressure-actuated cellular structures. \emph{Bioinspir. Biomim.} \textbf{7}, 016007. (doi:10.1088/1748-3182/7/1/016007)

	\bibitem{Pagitz2013} Pagitz M, Bold J. 2013. Shape-changing shell-like structures. \emph{Bioinspir. Biomim.} \textbf{8}, 016010. (doi:10.1088/1748-3182/8/1/016010)

	\bibitem{Freeman2009} Freeman E. 2009 Applications of biologically inspired membranes, M.S., University of Pittsburgh.

	\bibitem{Freeman2012} Freeman E. 2012 Harnessing Protein Transport for Engineering Applications: A Computational Study, Ph.D., University of Pittsburgh.

	\bibitem{Lenau2014} Lenau TA, Hesselberg T. 2014 Self-organisation and motion in plants. \emph{Proc. of SPIE} \textbf{9055}, 90550F(1-8). (doi:10.1117/12.2045155)

	\bibitem{Dicker} Dicker MPM, Rossiter JM, Bond IP, Weaver PM. 2014 Biomimetic photo-actuation: sensing, control and actuation in sun-tracking plants. \emph{Bioinspir. Biomim.} \textbf{9} 036015. (doi:10.1088/1748-3182/9/3/036015)

	\bibitem{YoshiokaCalvert} Yoshioka Y, Calvert P. 2002 Epoxy-based electroactive polymer gels. \emph{Sustain. Energy Rev.} \textbf{13} 1800-18. (doi:10.1007/BF02412145)

	\bibitem{Haswell2013} Monshausen GB, Haswell ES. 2013 A force of nature: molecular mechanisms of mechanoperception in plants. \emph{J. Exp. Bot.} \textbf{64}, 4663?-4680. (doi:10.1093/jxb/ert204)

	\bibitem{Campas2014} Campas O, Mammoto T, Hasso S, Sperling RA, O'Connell D, Bischof AG, Maas R, Weitz DA, Mahadevan L, Ingber DE. 2014 Quantifying cell-generated mechanical forces within living embryonic tissues. \emph{Nature Methods} \textbf{11}, 183-189. (doi:10.1038/nmeth.2761)

\end{thebibliography}
\end{document}